\documentclass{article}

\usepackage{geometry}
\usepackage{amsmath,amsthm,latexsym,amssymb,amsfonts}
\usepackage[parfill]{parskip}
\usepackage{accents}
\usepackage{graphicx,color}
\usepackage{hyperref}
\usepackage[utf8]{inputenc}
\usepackage[T1]{fontenc}
\usepackage{textcomp}

\def\equationautorefname~#1\null{eq.~(#1)\null
}

\hypersetup{
    unicode=false,          
    pdftoolbar=true,        
    pdfmenubar=true,        
    pdffitwindow=false, 
    pdfstartview={FitH},    
    pdftitle={Deformed Weitzenb\"ock connections, Teleparallel Gravity and Double field theory},    
    pdfauthor={Victor A. Penas},     
    pdfsubject={},   
    pdfnewwindow=true,      
    colorlinks=false,       
    linkcolor=blue,          
    citecolor=blue,        
    filecolor=blue,      
    urlcolor=blue           
    linkbordercolor={blue},
}

\newcommand{\dbar}[1]{\underaccent{\bar}{#1}}

\begin{document}

\begin{titlepage}
\begin{center}

\vskip 1.5cm

{\Large \bf
Deformed Weitzenb\"ock Connections and Double Field Theory
}

\vskip 1.5cm

{\bf  Victor A. Penas$^1$}

\vskip 30pt

{\em $^1$ \hskip -.1truecm G. F\'isica CAB-CNEA and CONICET,\\ Centro At\'omico Bariloche, Av. Bustillo 9500, Bariloche, Argentina}

\vskip 0.5cm

\small{victor.penas@cab.cnea.gov.ar}

\vskip 1cm

\end{center}

\vskip 0.5cm

\begin{center} {\bf ABSTRACT}\\[3ex]
\end{center}

We revisit the generalized connection of Double Field Theory. We implement a procedure that allow us to re-write the Double Field Theory equations of motion in terms of geometric quantities (like generalized torsion and non-metricity tensors) based on other connections rather than the usual generalized Levi-Civita connection and the generalized Riemann curvature. We define a generalized contorsion tensor and obtain, as a particular case, the Teleparallel equivalent of Double Field Theory. To do this, we first need to revisit generic connections in standard geometry written in terms of first-order derivatives of the vielbein in order to obtain equivalent theories to Einstein Gravity  (like for instance the Teleparallel gravity case). The results are then easily extrapolated to DFT.    

\end{titlepage}

\newpage
\setcounter{page}{1} \tableofcontents


\setcounter{page}{1} \numberwithin{equation}{section}

\section{Introduction}

In this work we revisit some geometric aspects of the generalized connection of Double Field Theory (DFT). DFT \cite{Siegel:1993xq,Hull:2009mi} (see \cite{Aldazabal:2013sca} for reviews) is a string-inspired field theory that describes the massless sector of string theory in a T-duality covariant way. Interestingly enough, many geometric notions similar to diffeomorphisms, covariant derivatives (connections) and curvature tensors arise in a natural and generalized way, transcending the usual framework of differential geometry. Its geometric structure, in fact, is closely related to that of generalized geometry and Courant algebroids  \cite{Hitchin:2004ut,Coimbra:2011nw, Grana:2008yw,Chatzistavrakidis:2018ztm,Cederwall:2014kxa,Vaisman:2012ke}. 

However, in contrast to General Relativity (GR), the DFT version of the Levi-Civita connection has not all of its components fully determined in terms of the physical fields, usually taken to be the generalized metric and dilaton, but only some projections of the connection are. This also implies that the DFT curvature tensor is not fully determined in terms of the physical fields \cite{Siegel:1993xq,Hohm:2011si}. There were other attempts to try to fully determine the DFT connection. For instance,  in \cite{Jeon:2010rw} a DFT-connection was fully determined by demanding usual compatibility criteria plus an extra constraint (the connection was required to live in the kernel of a particular six-index tensor). The result was a semi-covariant derivative, i.e. only projected components of it transform well under generalized coordinate transformations. In \cite{Berman:2013uda}, the Weitzenb\"ock connection was considered rather than the generalized Levi-Civita.  The resulting theory, based on this connection, reproduces the DFT action (up to a boundary term) and the dynamics is based on the (generalized) torsion. However, a reformulation in terms of other geometric quantities (like the contorsion tensor) and its analogy with teleparallel gravity (TG) was not analyzed. In the context of DFT on group manifolds \cite{Blumenhagen:2014gva,Blumenhagen:2015zma, Hassler:2016srl} a fully determined covariant derivative was found that has non-trivial curvature and non-trivial torsion. But this covariant derivative is not covariant with respect to generalized diffeomorphisms. Instead it is covariant with respect to normal 2D-diffeomorphims\footnote{We thank Falk Hassler for pointing this out.}. More recently, in \cite{Freidel:2018tkj} a fully determined DFT connection is proposed by extending the geometry of DFT with a para-Hermitian structure, resulting in a Born Geometry. 

In  this work we return to the issue of finding fully determined connections in DFT. We propose a procedure that outlines why it is possible to consider fully determined connections (besides the generalized Levi-Civita and Weitzenb\"ock connections) in DFT and to understand their geometric structures. These connections will reproduce theories equivalent to DFT and, in particular, we obtain the teleparallel equivalent of DFT. We can interpret these new connections as deformed versions of the Weitzenb\"ock connection. This procedure  will also allow us, in a natural way, to reinterpret and determine the undetermined parts of the generalized Levi-Civita connection. The gauge and global symmetries will play a crucial role and to understand this procedure we first need to briefly review the teleparallel equivalent of GR.

Teleparallel Gravity (TG or TEGR) is an alternative formulation to General Relativity (GR)\cite{firstrefs,Hayashi:1979qx,Arcos:2005ec}. It is usually based on parallelizable manifolds equipped with the Weitzenb\"ock connection (the connection of the parallelization) rather than the Levi-Civita connection. One of the key aspects of the theory is that the Weitzenb\"ock connection furnishes a null Riemann curvature tensor, and since this connection is metric-compatible, the dynamics of the theory is based only on the torsion. The Weitzenb\"ock connection is of the form $W_{\mu \nu}{}^{\rho}= \partial_{\mu} e^{a}{}_{\nu}  e_{a}{}^{\rho}$, where $\mu$ and $a$ are curved and flat indices respectively, and $e_{a}{}^{\mu}$ is the vielbein. The vielbein is treated as the fundamental field of the theory and the metric is interpreted as a byproduct coming from it.      
The name of TG comes from the fact that there is a notion of absolute parallelism between vectors located at different space-time points. Indeed, one way to see this is to notice that the vielbein field is covariantly constant with respect to the  Weitzenb\"ock connection in a coordinate basis. This means that we can compare the flat components of a vector field defined at different space-time points just like we would do in a global basis of flat space-time. The incorporation of sources have also been studied (see \cite{deAndrade:1997gka}) in the TG case. As a remark, TG has been found to have applications in cosmology by studying theories called $f(T)$-Theories \cite{Ferraro:2006jd, Bengochea:2008gz, Linder:2010py, Bamba:2010wb, Li:2011wu, Cai:2015emx, Paliathanasis:2016vsw}, where $T$ stands for scalar combinations built out of the torsion. 

The key observation to our work is that in TG the local Lorentz symmetry group is generically broken\footnote{ In $f(T)$-theories at most only a remnant subgroup of it survives \cite{Li:2010cg,Sotiriou:2010mv,Ferraro:2014owa}.}. It is easy to see that space-time tensors built out of the Weitzenb\"ock connection (for instance the torsion $T_{\mu\nu}{}^{\rho}$) will not always transform as scalars under local Lorentz transformations. In fact, the Weitzenb\"ock connection itself, in a coordinate basis, does not transform like a scalar under local Lorentz transformations but under global ones (i.e. with constant Lorentz parameters). Global transformations bring the immediate worry about extra degrees of freedom coming from the vielbein, since in $D=4$, six out of the 16 components of the vielbein are gauged away by local Lorentz transformations. However, since the TG action turns out to be equal to the Einstein-Hilbert action, up to a boundary term, the equations of motion of TG are equivalent to the Einstein equations and thus the theory enjoys local Lorentz symmetry.

Motivated by this global-local mechanism of TG, and inspired by the necessity of DFT to consider connections (and curvatures) determined in terms of the physical fields of the theory, we will try to obtain equivalent theories to DFT defined by fully determined connections. Although these connections will transform under double Lorentz transformations with constant parameters, the resulting theories will possess double local Lorentz symmetry (that is, under parameters depending on the coordinates). But before doing so, we first need to apply this procedure to GR. That is, our starting point will be to first understand the most general connection written in terms of first-order derivatives of the vielbein\footnote{We will not consider terms that could involve the epsilon tensor $\epsilon_{\mu\nu\rho\sigma}$.} in standard geometry: 
\begin{align}\nonumber
\Gamma_{\mu \nu}{}^{\rho}= W_{\mu\nu}{}^{\rho} - {a}_{1} \overset{(W)}{T}_{\nu \mu}\,^{\rho} + {b}_{1} \overset{(W)}{T}{}^{\rho}\,_{\mu \nu} + {b}_{2} \overset{(W)}{T}{}^{\rho}\,_{\nu \mu}\\\label{GammaConnectioncurvedsix}
- {c}_{1} g_{\mu \nu} \overset{(W)}{T}_{\sigma}\,^{\rho \sigma} - {d}_{1} {\delta}_{\mu}\,^{\rho} \overset{(W)}{T}_{\nu}\,^{\sigma}\,_{\sigma} - {d}_{2} {\delta}_{\nu}\,^{\rho} \overset{(W)}{T}_{\mu}\,^{\sigma}\,_{\sigma}.
\end{align}
Where $\overset{(W)}{T}_{\mu\nu}{}^{\rho}=2W_{[\mu\nu]}{}^{\rho}$ is the torsion of the Weitzenb\"ock connection\footnote{We are using the following conventions on indices: $(a b)=\frac{1}{2}\Big(a b + b a\Big)$ and $[a b]=\frac{1}{2}\Big(a b - b a\Big)$. In section (\ref{eqforCoeff}) we will denote quantities associated with the Weitzenb\"ock connection with a $(4)$ as superscript, e.g. $\overset{(W)}{T}_{\mu\nu}{}^{\rho}=\overset{(4)}{T}_{\mu\nu}{}^{\rho}$.} and $g_{\mu\nu}$ is the spacetime metric. The connection (\ref{GammaConnectioncurvedsix}) is parametrized by six real parameters and it does not have any symmetry requirement imposed on its indices so it will yield, generically, non-vanishing curvature, torsion and non-metricity. In a coordinate basis, this connection transforms in  a proper way under general coordinate transformations and is a scalar under \emph{constant} Lorentz transformations. This means that for generic coefficients, the local Lorentz symmetry is broken. In DFT, we will repeat the same analysis as in standard geometry and we will also consider the most general connection written in terms of first-order derivatives acting on the generalized vielbein (see subsection (\ref{DFTsection2exampotherconn})). This DFT-connection is a well-defined connection under generalized coordinate transformations and a scalar under double Lorentz transformations with constant parameters. A key role will be play by the generalized Weitzenb\"ock connection and the introduction of generalized contorsion and non-metricity tensors.

The paper is organized as follows. In section (\ref{Notation}) we will introduce the notation used in the paper and show how to obtain different actions equivalent to the Einstein-Hilbert action for connections with generic features. The idea will be to set to zero a generic scalar curvature and extract from it a tensor quantity (based on generic torsion and non-metricity) that yields the same equations of motion given by the usual Ricci scalar (based on the Levi-Civita). As far as we know, these connections have not been explicitly classified in a systematic way by imposing the vanishing of a generic scalar curvature, neither in GR\footnote{In \cite{Itin:2004wv} a generic contorsion tensor  and non-metricity tensor built from a 6-parametric connection (similar to (\ref{GammaConnectioncurvedsix})) constructed from the coframe were considered. Although there are several similarities with our results, an analysis of the parameters extracted from requiring the vanishing of a generic scalar curvature has not been done. This is a general procedure that allows us to classify and write down explicit expressions for the parameters of the possible connections, and also to derive equations of motion. This procedure is easily extrapolated to DFT. We also want to stress that, we do not have extra degrees of freedom in the dynamical equations. In this sense, we differ with respect to the Metric-Affine Gravity theories where they usually consider extra degrees of freedom \cite{Hehl:1994ue}. See \cite{Obukhov:2002tm} also for a treatment of TG in metric affine space-times.} nor in DFT. In subsection (\ref{eqforCoeff}) we will find constraints on the parameters of (\ref{GammaConnectioncurvedsix}). The parameters that satisfy these constraints will give the desired connections that allow for actions equivalent, up to a boundary term, to the Einstein-Hilbert action and we will analyze particular values of these parameters. In subsection (\ref{eqofmotionsection}) we will obtain the equations of motion for these theories based on the new connections.
In section (\ref{DFTsection2}) we proceed to implement the same analysis as in the previous section and remark similarities and differences. We analyze the components of the generalized connection without imposing any compatibility constraint. In subsection (\ref{DFTsection2fullconnection}) we present the general decomposition of the generalized connection and define a generalized contorsion tensor and a L-tensor (made up of the non-metricity DFT tensor). In subsection (\ref{DFTsection2genriemanntensor})  we compute the decomposition of the generalized Riemann tensor which allows us to compute the equations of motion, derived from the new connection, in subsection (\ref{DFTsection2EoM}). In subsection (\ref{DFTsection2undetpartconn}) we talk about how the undetermined parts of the connection get naturally determined. In subsection  (\ref{DFTsection2TDFT}) we introduce the Teleparallel Equivalent of Double Field Theory and in (\ref{DFTsection2exampotherconn}) we show what structures arise for other connections that present generically non-vanishing curvature and torsion.
 In section (\ref{conclusions}) we summarize our work. 
 
 It is not often known that GR admits the possibility to be described by other connections rather than the usual ones mentioned before and we will show that in DFT this is also the case. The DFT connections will turn out to be naturally described (and fully determined) in a vielbein formulation by defining notions of contorsion and non-metricity. More importantly, these connections will obey the same global-local mechanism of TG theories, and, in particular, we will see that the generalized Levi-Civita connection will also naturally do it. We believe these results are a step forward to better understand and clarify the mathematical structures of the DFT connections and also for future applications in $\alpha'$-corrections as mentioned in the Summary section.

\section{Connections in General Relativity}\label{Notation}

In this section we will work in $D=4$ spacetime dimensions, $\mu,\nu...$ are curved indices and $a,b...$ denote flat indices. Our fundamental field is the vielbein $e_{a}{}^{\mu}$ and we define the metric as a byproduct $g_{\mu \nu}=\eta_{a b} e^{a}{}_{\mu} e^{b}{}_{\nu}$. Here $\eta_{a b}$ is the constant Minkowski metric. 
The vielbein satisfies $e_{a}{}^{\mu} e^{b}{}_{\mu} = \delta_{a}{}^{b}$ and $e^{a}{}_{\mu} e_{a}{}^{\nu} = \delta_{\mu}{}^{\nu}$ where we raise and lower curved and flat indices with the metric $g_{\mu \nu}$ and $\eta_{a b}$ respectively. We define the flat derivative $D_{a}=e_{a}{}^{\mu}\partial_{\mu}$. It is useful to introduce the following quantities:
\begin{equation}
\Omega_{a b}{}^{c}=D_{a} e_{b}{}^{\rho} e^{c}{}_{\rho},~~~~\tau_{a b}{}^{c}=2\Omega_{[a b]}{}^{c}.
\end{equation}
Here, $\tau_{a b}{}^{c}$ are the anholonomy coefficients. The Riemann curvature tensor is defined as follows:
\begin{equation}\label{RiemannCurved}
R_{\mu \nu \rho}{}^{\sigma}=\partial_{\mu} \Gamma_{\nu \rho}{}^{\sigma} - \partial_{\nu} \Gamma_{\mu \rho}{}^{\sigma} - \Gamma_{\mu \rho}{}^{\lambda} \Gamma_{\nu \lambda}{}^{\sigma} + \Gamma_{\nu \rho}{}^{\lambda} \Gamma_{\mu \lambda}{}^{\sigma}. 
\end{equation}
For generic affine connections, the Riemann tensor is only antisymmetric in the first two indices. In a metric-affine spacetime the affine connection admits a general decomposition of the form:
\begin{equation}\label{eqDecompOfGamma}
\Gamma_{\mu \nu}{}^{\sigma}= \mbox{\scriptsize{$\left\{\begin{array}{@{}c@{}} \sigma \\ \mu ~\nu\end{array}\right\}$}} + K_{\mu \nu}{}^{\sigma} + L_{\mu \nu}{}^{\sigma},
\end{equation}
where the Christoffel symbols, contorsion tensor and the $L_{\mu\nu}{}^{\rho}$ tensor are given respectively by:
\begin{equation}
\mbox{\scriptsize{$\left\{\begin{array}{@{}c@{}} \sigma \\ \mu ~\nu\end{array}\right\}$}}=\frac{1}{2} g^{\tau \sigma}\left(\partial_{\mu} g_{\nu \tau} - \partial_{\tau} g_{\mu \nu} + \partial_{\nu} g_{\tau \mu}\right), 
\end{equation}
\begin{equation}
K_{\mu \nu}{}^{\sigma}=-\frac{1}{2}g^{\tau \sigma}\left(T_{\mu \tau}{}^{\rho} g_{\rho \nu} - T_{\tau \nu}{}^{\rho} g_{\rho \mu} - T_{\mu \nu}{}^{\rho} g_{\rho \tau}\right),
\end{equation}
\begin{equation}
L_{\mu \nu}{}^{\sigma}=-\frac{1}{2} g^{\tau \sigma}\left(\nabla_{\mu} g_{\nu \tau} - \nabla_{\tau} g_{\mu \nu} + \nabla_{\nu} g_{\tau \mu}\right). 
\end{equation}
Here, the torsion is $T_{\mu\nu}{}^{\sigma}=2\Gamma_{[\mu\nu]}{}^{\sigma}$ and the covariant derivative is defined as $\nabla_{\mu}V_{\nu}=\partial_{\mu} V_{\nu} - \Gamma_{\mu \nu}{}^{\rho} V_{\rho}$. The combination $Q_{\mu\nu\rho}=\nabla_{\mu}g_{\nu\rho}$ is known as the non-metricity tensor. When we plug the above decomposition in (\ref{RiemannCurved}) we obtain:
\begin{align}\nonumber
R_{\mu\nu\rho}{}^{\sigma}(\Gamma)&=R_{\mu\nu\rho}{}^{\sigma}(\{\}) + \overset{(\{\})}{T}_{\mu \nu}{}^{\lambda}C_{\lambda\rho}{}^{\sigma} \\\label{eqDecompOfRiemann}
&~~~~+ 2\overset{(\{\})}{\nabla}_{[\mu} C_{\nu]\rho}{}^{\sigma} + 2C_{[\mu|\lambda|}{}^{\sigma}C_{\nu]\rho}{}^{\lambda}\, ,
\end{align}
where $C_{\mu\nu}{}^{\sigma}=K_{\mu\nu}{}^{\sigma} + L_{\mu\nu}{}^{\sigma}$ and $\overset{(\{\})}{\nabla}_{\mu}$ is referred to the Levi-Civita connection. The Ricci tensor and Ricci scalar are defined as
\begin{equation}
R_{\mu \nu}=R_{\mu \rho \nu}{}^{\rho}, ~~~~R=g^{\mu \nu}R_{\mu \rho \nu}{}^{\rho}.
\end{equation}
The decomposition (\ref{eqDecompOfRiemann}) on the Ricci scalar yields:
\begin{align}\nonumber
R(\Gamma)&=R(\{\}) + \overset{(\{\})}{\nabla}_{\mu}C_{\sigma}{}^{\mu\sigma} - \overset{(\{\})}{\nabla}_{\sigma}C_{\mu}{}^{\mu\sigma}\\\label{Rscalardecomposed}
&~~~~~~~~~~~~~~~~ + C_{\mu\lambda}{}^{\sigma} C_{\sigma}{}^{\mu\lambda} - C_{\sigma\lambda}{}^{\sigma}C_{\mu}{}^{\mu\lambda}.
\end{align} 
The Levi-Civita connection is torsionless and metric-compatible, which means that $\overset{(\{\})}{T}_{\mu\nu}{}^{\sigma}=0$ and the metric can pass through the covariant derivative.

The relation between the components of the affine connection written in a holonomic basis and an anholonomic one is of the form:
\begin{equation}\label{affineandgaugeconn}
\Gamma_{\mu \nu}{}^{\rho}=W_{\mu \nu}{}^{\rho} + e ^{b}{}_{\nu} e_{c}{}^{\rho} w_{\mu b}{}^{c}.
\end{equation} 
For instance, if $\Gamma_{\mu\nu}{}^{\rho}$ is the Levi-Civita connection then $w_{\mu a}{}^{b}$ is the usual spin connection (\ref{SpinConnLevCivTau}). For simplicity, we will refer to $w_{\mu a}{}^{b}$ as a gauge connection. By inserting (\ref{GammaConnectioncurvedsix}) in (\ref{affineandgaugeconn}) and using $W_{\mu \nu}{}^{\rho}=-e^{a}{}_{\mu} e^{b}{}_{\nu} e_{c}{}^{\rho} \Omega_{a b}{}^{c}$ the gauge connection gets the following form: 
\begin{align}\nonumber
w_{a b}{}^{c} = & \, {a}_{1} {\tau}_{b a}\,^{c} - {b}_{1} {\tau}^{c}\,_{a b} - {b}_{2} {\tau}^{c}\,_{b a} + {c}_{1} {\eta}_{a b} {\tau}_{d}\,^{c d}\\\label{SpinConnectionTau}
& + {d}_{1} {\delta}_{a}\,^{c} {\tau}_{b}\,^{d}\,_{d} + {d}_{2} {\delta}_{b}\,^{c} {\tau}_{a}\,^{d}\,_{d}.
\end{align}
It is important to stress that the above gauge connection (with all flat indices) transforms as a scalar under diffeomorphisms. However, for generic coefficients, it does not transform like a gauge connection under local gauge transformations, but only global (constant) ones. In fact, if we force this connection to transform under local Lorentz transformations we see that the only possibility is with $a_{1}=b_{1}=b_{2}=-1/2$,  $c_{1}=d_{1}=d_{2}=0$. This is precisely the coefficients for the Levi-Civita spin connection:
\begin{align}\label{SpinConnLevCivTau}
\overset{(\{\})}{w}_{a b}{}^{c}= \frac{1}{2}\, {\tau}_{a b}\,^{c} + \frac{1}{2}\, {\tau}^{c}\,_{a b} + \frac{1}{2}\, {\tau}^{c}\,_{b a}\,.
\end{align}
In planar indices, the Riemman tensor (\ref{RiemannCurved}) takes the form:
\begin{align}\nonumber
R_{a b c}{}^{d} = & {D}_{a}{{w}_{b c}\,^{d}}\,  - {D}_{b}{{w}_{a c}\,^{d}}\,  - {w}_{a c}\,^{e} {w}_{b e}\,^{d} \\\label{RiemannFlat}
&+ {w}_{b c}\,^{e} {w}_{a e}\,^{d} - {\tau}_{a b}\,^{e} {w}_{e c}\,^{d}.
\end{align}
Similarly as before, the Ricci tensor and the Ricci scalar are of the form $R_{a b}=R_{a c b}{}^{c}$ and  $R=\eta^{a b}R_{a b}$ respectively.

The idea is to use the connection (\ref{SpinConnectionTau}) (or equivalently (\ref{GammaConnectioncurvedsix})) and find conditions on the coefficients ($a_{1}$, $b_{1}$, $b_{2}$, $c_{1}$, $d_{1}$, $d_{2}$) such that the Ricci scalar for this connection vanishes. In this way, we obtain an equality between the usual Einstein-Hilbert action and an action with $C$-terms (up to a boundary term). Indeed, consider the Einstein-Hilbert action with the scalar curvature (\ref{Rscalardecomposed}):
\begin{align}\nonumber
&\int dx^4\sqrt{-g}R(\Gamma)=  \int dx^4 \sqrt{-g}\Big(R(\{\}) \\\label{FundamentalEquationone}
& ~~~~~~~~~~~~~~~~~~~~~+ C_{\mu\lambda}{}^{\sigma}C_{\sigma}{}^{\mu\lambda} - C_{\sigma\lambda}{}^{\sigma}C_{\mu}{}^{\mu\lambda}\Big)\,.
\end{align}
We have dropped out the covariant derivative terms since they form a total derivative.
By setting $R(\Gamma)=0$ in (\ref{FundamentalEquationone}) we simply obtain:
\begin{align}\nonumber
&\int dx^{4} \sqrt{-g}R(\{\}) = \\\label{FundamentalEquationtwo}
&~~~~~=- \int dx^{4} \sqrt{-g}\left(C_{\mu\lambda}{}^{\sigma}C_{\sigma}{}^{\mu\lambda} - C_{\sigma\lambda}{}^{\sigma}C_{\mu}{}^{\mu\lambda}\right).
\end{align}

We stress again that the above equality should be understood up to a boundary term. In the second line of (\ref{FundamentalEquationtwo}), the tensor $C_{\mu\nu}{}^{\rho}$ depends on the general connection $\Gamma$ (\ref{GammaConnectioncurvedsix}). This means that  connections that furnish a null Ricci scalar can be used to yield an action equivalent to the Einstein-Hilbert action by using its torsion and/or non-metricity. This is indeed the case for the usual TG, where the right hand side of (\ref{FundamentalEquationtwo}) reproduces exactly the teleparallel action with $C_{\mu\nu}{}^{\rho}(W)=K_{\mu\nu}{}^{\rho}(W)$ (see section \ref{subunoEOMsec}).

The reader might be worried about the fact that we are considering global Lorentz transformations. But all of the connections we will obtain reproduce theories which possess local gauge transformations as established by the action (\ref{FundamentalEquationtwo}), reducing in this way the number of degrees of freedom of the vielbein. This is the same mechanism as in teleparallel gravity\footnote{At least for the usual teleparallel gravity which is equivalent to Einstein gravity (TEGR). As mentioned in the introduction, the Lorentz transformations in $f(T)$-gravities are more subtle.} where the affine connection is chosen to be the Weitzenb\"ock connection $W_{\mu\nu}{}^{\rho}$ which is invariant only under Lorentz transformations with constant parameters but the resulting theory turns out to have local Lorentz symmetry.   

For simplicity we will mostly work with flat indices.  
The usual Einstein-Hilbert action with Levi-Civita connection after a partial integration is of the form: 
\begin{align}\nonumber
\int dx^{4} \sqrt{-g} R(\{\})= & \int dx^{4} \sqrt{-g} \Big(-\tau_{a c}{}^{c} \tau^{a b}{}_b\, \\
&+ \frac{1}{2}\, {\tau}^{a}\,_{b}\,^{c} {\tau}_{a c}\,^{b} + \frac{1}{4}\, {\tau}^{a b}\,_{c} {\tau}_{a b}\,^{c} \Big),
\end{align}
where we have used (\ref{RiemannFlat}) and (\ref{SpinConnLevCivTau}). Equation (\ref{FundamentalEquationtwo}) then gets the form:

\begin{align}\nonumber
\int dx^{4} \sqrt{-g} \Big(&-\tau_{a c}{}^{c} \tau^{a b}{}_b\,  + \frac{1}{2}\, {\tau}^{a}\,_{b}\,^{c} {\tau}_{a c}\,^{b}  + \frac{1}{4}\, {\tau}^{a b}\,_{c} {\tau}_{a b}\,^{c} \Big)\\ \label{FundamentalEquationthree}
&~= - \int dx^{4} \sqrt{-g}\Big(C_{a b}{}^{c}C_{c}{}^{a b} - C_{c b}{}^{c}C_{a}{}^{a b}\Big).
\end{align}

\subsection{Equations for coefficients}\label{eqforCoeff}

As mentioned before, we would like to find conditions on the coefficients of the general connection (\ref{SpinConnectionTau}) such that the Ricci scalar of this connection vanishes. We plug (\ref{SpinConnectionTau}) in the Riemann tensor (\ref{RiemannFlat}) and contract indices to obtain the  scalar curvature. The result is the following:
\begin{align}\nonumber
R(w)=& {D}^{a}{{\tau}_{a}\,^{b}\,_{b}}\,  \Big({a}_{1} + 2\, {b}_{2} + 3\, {c}_{1} + 3\, {d}_{1} + {b}_{1}\Big) \\\nonumber
&+ {\tau}^{a b c} {\tau}_{a b c} \Big({a}_{1} {b}_{1} - {b}_{1} {b}_{2} - {a}_{1} {b}_{2} - {b}_{2}\Big)\\\nonumber
& + {\tau}^{a b c} {\tau}_{a c b} \Big( - {a}_{1} {a}_{1} + {a}_{1} {b}_{2} - {a}_{1} {b}_{1} - {b}_{1} {b}_{1} \\\nonumber
&~~~~~~~~~~~~~~~~~ + {b}_{1} {b}_{2} - {b}_{2} {b}_{2} - {a}_{1} - {b}_{1}\Big)\\\nonumber
&  + {\tau}_{b}\,^{c}\,_{c} {\tau}^{a b}\,_{a} \Big( - {a}_{1} {b}_{1} - {b}_{1} {b}_{2} + 2\, {b}_{1} {c}_{1} - 4\, {b}_{1} {d}_{1} \\\nonumber
& ~~~~~~~~~~~~~~~~~~ - {a}_{1} {b}_{2} - {b}_{2} {b}_{2} - 2\, {b}_{2} {c}_{1} - 2\, {b}_{2} {d}_{1}\\\nonumber
&~~~~~~~~~~~~~~~~~~ - 4\, {a}_{1} {c}_{1} + 3\, {c}_{1} {c}_{1} - 12\, {c}_{1} {d}_{1}\\\label{RicciScalarSpinConnTau}
&~~~~~~~~~~~~~~~~~~ + 2\, {a}_{1} {d}_{1} + 3\, {d}_{1} {d}_{1} + {c}_{1} + {d}_{1}\Big);
\end{align}

All of the terms are independent of each other so in order to obtain a null scalar curvature we must set each parentheses to zero. 
This defines a set of four quadratic equations for five variables ($a_{1}$, $b_{1}$, $b_{2}$, $c_{1}$, $d_{1}$). Note that the coefficient $d_{2}$ has dropped out from (\ref{RicciScalarSpinConnTau}) but does not drop out from the Riemann tensor nor the Ricci tensor. The reason is the following. It is known that a Riemann tensor with a generic connection is invariant under:
\begin{equation}\label{gaugetransconncurvature}
\Gamma_{\mu\nu}{}^{\rho}\rightarrow \Gamma_{\mu\nu}{}^{\rho} + \delta_{\nu}{}^{\rho}\partial_{\mu}\phi,
\end{equation}
where $\phi$ is a scalar. However, the Ricci scalar built from this generic Riemann tensor is invariant under a relaxed version of the above equation: 
\begin{equation}\label{gaugetransfRicciScalar}
\Gamma_{\mu\nu}{}^{\rho}\rightarrow \Gamma_{\mu\nu}{}^{\rho} +A_{\mu}\delta_{\nu}{}^{\rho}.
\end{equation}
This is due to a contraction between the metric and $\partial_{[\mu}A_{\nu]}$. We can see that the term with coefficient $d_{2}$ in  (\ref{GammaConnectioncurvedsix}) has exactly the same form as in (\ref{gaugetransfRicciScalar}). We will come back to this point later.
 
We might have set to zero the curvature tensor or the Ricci tensor in order to obtain equations for the coefficients. Nevertheless, we analyze the vanishing of the scalar curvature because the solutions for this one includes the cases for a null curvature tensor and Ricci tensor. Therefore, teleparallel gravity must show up as a particular solution to our set of four quadratic equations derived from (\ref{RicciScalarSpinConnTau}).\\ 

Our set of four equations have an infinite amount of solutions yielding both torsion and non-metricity (and, in general, a non-vanishing curvature). To simplify the analysis we consider particular cases in the following subsections.

\subsection{Metric-Compatible case}\label{eqforCoeffMCcase}

In this subsection we restrict to the case of metric-compatible connections, i.e.:
\begin{equation}
D_a \eta_{a b} - w_{a b}{}^{e}\eta_{e c} - w_{a c}{}^{e}\eta_{b e} = 0 ~~\Rightarrow ~~ w_{a (b c)} = 0\,. 
\end{equation}
The above condition implies 
\begin{equation}\label{condmetriccompat}
a_{1}=b_{1},~~c_{1}=d_{1},~~d_{2}=0,
\end{equation}
The set of four equations derived from (\ref{RicciScalarSpinConnTau}) together with conditions (\ref{condmetriccompat}) yield only four solutions which will be referred to as cases 1 to 4. These are respectively:

\begin{equation}\label{QzeroSolOne}
a_{1}=-1,~~ b_{2}=-1, ~~c_{1}= 2/3,
\end{equation}
\begin{equation}\label{QzeroSolTwo}
a_{1}=-2/3,~~ b_{2}=-4/3,~~ c_{1}= 2/3,
\end{equation}
\begin{equation}\label{QzeroSolThree}
a_{1}=-1/3,~~ b_{2}=1/3,~~ c_{1}=0,
\end{equation}
\begin{equation}\label{QzeroSolFour}
a_{1}=0,~~ b_{2}=0,~~ c_{1}=0,
\end{equation}
Replacing cases 1 to 4 in the general connection (\ref{GammaConnectioncurvedsix}) we obtain the following connections, respectively:  
\begin{align}\nonumber
\overset{(1)}{\Gamma}_{\mu \nu}{}^{\rho}= W_{\mu\nu}{}^{\rho} &+ \overset{(W)}{T}_{\nu \mu}\,^{\rho} - \overset{(W)}{T}{}^{\rho}\,_{\mu \nu} - \overset{(W)}{T}{}^{\rho}\,_{\nu \mu}\\\label{QzeroSolOneConn}
&- \frac{2}{3} g_{\mu \nu} \overset{(W)}{T}_{\sigma}\,^{\rho \sigma} - \frac{2}{3} {\delta}_{\mu}\,^{\rho} \overset{(W)}{T}_{\nu}\,^{\sigma}\,_{\sigma}.
\end{align}
\begin{align}\nonumber
\overset{(2)}{\Gamma}_{\mu \nu}{}^{\rho}= W_{\mu\nu}{}^{\rho} &+\frac{2}{3} \overset{(W)}{T}_{\nu \mu}\,^{\rho} -\frac{2}{3} \overset{(W)}{T}{}^{\rho}\,_{\mu \nu} -\frac{4}{3} \overset{(W)}{T}{}^{\rho}\,_{\nu \mu}\\\label{QzeroSolTwoConn}
&- \frac{2}{3} g_{\mu \nu} \overset{(W)}{T}_{\sigma}\,^{\rho \sigma} - \frac{2}{3} {\delta}_{\mu}\,^{\rho} \overset{(W)}{T}_{\nu}\,^{\sigma}\,_{\sigma}.
\end{align}
\begin{align}\nonumber
\overset{(3)}{\Gamma}_{\mu \nu}{}^{\rho}= W_{\mu\nu}{}^{\rho} &+\frac{1}{3} \overset{(W)}{T}_{\nu \mu}\,^{\rho} -\frac{1}{3} \overset{(W)}{T}{}^{\rho}\,_{\mu \nu} + \frac{1}{3} \overset{(W)}{T}{}^{\rho}\,_{\nu \mu}.\\\label{QzeroSolThreeConn}
\end{align}
\begin{align}\label{QzeroSolFourConn}
\overset{(4)}{\Gamma}_{\mu \nu}{}^{\rho}= W_{\mu\nu}{}^{\rho}.
\end{align}
Equivalently, we use (\ref{SpinConnectionTau}) with the above coefficients.  As expected, one of the solutions (eq. (\ref{QzeroSolFourConn})) is the Weitzeb\"ock connection.

The above connections (\ref{QzeroSolOneConn})-(\ref{QzeroSolFourConn}) are metric compatible (the non-metricity tensor vanishes) and present non-vanishing torsion. Thus only the contorsion tensor $K_{a b}{}^{c}$ will contribute to (\ref{FundamentalEquationthree}).
The contorsion tensor for the metric-compatible case is equal to: 
\begin{equation}\label{defcontorsionMCcase}
K_{a b}{}^c=C_{a b}{}^c=w_{a b}{}^c - \overset{(\{\})}{w}_{a b}{}^c.
\end{equation}
It is straightforward to check that (\ref{FundamentalEquationthree}) is satisfied for the four cases when plugging (\ref{defcontorsionMCcase}) in (\ref{FundamentalEquationthree}). That is:
\begin{align}\nonumber
& \overset{(i)}{K}_{a b}\,^{e} \overset{(i)}{K}_{e c}\,^{b} {s}^{c a} - \overset{(i)}{K}_{a b}\,^{a} \overset{(i)}{K}_{e c}\,^{b} {s}^{c e}=\\  &~~~~~~=- \frac{1}{2}\, {\tau}^{a b c} {\tau}_{a c b} - \frac{1}{4}\, {\tau}^{a b c} {\tau}_{a b c} + {\tau}^{b a}\,_{a} {\tau}_{b}\,^{c}\,_{c},
\end{align}
with $i=1,\cdots,4$ labeling cases 1 to 4.
We want to remark that the above connections not only yield non-vanishing torsion but also non-vanishing  curvature tensor and Ricci tensor. The only exception being case 4, i.e. the Weitzenb\"ock case $\overset{(4)}{\Gamma}_{\mu\nu}{}^{\rho}=W_{\mu\nu}{}^{\rho}$ (only has non-vanishing torsion).
\subsection{Non-metricity case}\label{eqforCoeffNMcase}
We begin studying the simplest case of torsionless non-metric connection. The condition is:
\begin{equation}
T_{\mu \nu}{}^\sigma = 0 \Rightarrow \tau_{a b}{}^c = w_{a b}{}^c - w_{b a}{}^c,
\end{equation}
which implies that the coefficients must satisfy:
\begin{equation}\label{nonmetrtorsionlesscase}
\begin{split}
1+2a_1=0,\\
b_1-b_2=0,\\
d_2-d_1=0.
\end{split}
\end{equation}
However, equations (\ref{nonmetrtorsionlesscase}) together with the set of four quadratic equations derived from (\ref{RicciScalarSpinConnTau}) yield no solution. 
There have been other attempts for constructing equivalent theories to EG based purely on non-metricity. For instance in \cite{Nester:1998mp}  a vanishing $\Gamma_{\mu\nu}{}^{\rho}$ was considered, allowing the theory to be described only in terms of the non-metricity tensor $Q_{\mu\nu\rho}=\partial_{\mu}g_{\nu\rho}$. This is equivalent as choosing $w_{a b}{}^{c}=e_{a}{}^{\mu}e_{b}{}^{\nu}e^{c}{}_{\rho}W_{\mu\nu}{}^{\rho}$. We believe that if we had allowed an extra arbitrary constant factor in front of $W_{\mu\nu}{}^{\rho}$ in (\ref{GammaConnectioncurvedsix}) then our procedure would have also yielded this gauge connection as a particular result. However, the transformation properties of this connection  under diffeomorphisms would have been compromised.

As said before there are infinite solutions to our set of four equations presenting, in a generic way, non-metricity and torsion. Just to analyze a particular case we study the condition for a Weyl's space\footnote{In \cite{Obukhov:1982zn} a Weyl vector has also been considered. However, we are not considering this vector to be an independent degree of freedom since it is built from the vielbein.}: 
\begin{equation}
\nabla_{\mu} g_{\nu\rho} = - 2A_{\mu}g_{\nu\rho} ~~\Rightarrow ~~w_{a (b c)} = A_{a} \eta_{b c}.  
\end{equation}  
Where $A_{\mu}$ is a one-form field (Weyl's vector field). Since we want a connection only in terms of derivatives of the vielbein the only possibility is $A_{a}=\alpha \tau_{a}{}^{d}{}_{d}$ for some real constant $\alpha$. This implies the conditions
 \begin{equation}\label{weylconditioncoeff}
 	a_{1}=b_{1}, ~c_{1}=d_{1}, ~\alpha=d_{2}. 
 \end{equation}
 These  conditions on the coefficients are the same as the previous case (\ref{condmetriccompat}) except that we have a non-trivial condition on $d_{2}$. Since $d_{2}$ does not show up in (\ref{RicciScalarSpinConnTau}), the solutions to the quadratic equations plus (\ref{weylconditioncoeff}) are the same as the previous case but with an extra term, i.e.:
\begin{equation}\label{gammaweylspace}
\overset{(i)}{\widetilde{\Gamma}}_{\mu\nu}{}^{\rho}=\overset{(i)}{\Gamma}_{\mu\nu}{}^{\rho} -\alpha \overset{(W)}{T}_{\mu}{}^{\sigma}{}_{\sigma}\delta_{\nu}{}^{\rho},
\end{equation}

with $i=1,\cdots,4$. Although these connections are different from (\ref{QzeroSolOneConn})-(\ref{QzeroSolFourConn}) due to the last term in (\ref{gammaweylspace}), we must notice that this extra term is just of the form (\ref{gaugetransfRicciScalar}) which leaves the Ricci scalar invariant. More explicitly: $A_{\mu}\delta_{\nu}{}^{\rho}=-\alpha \overset{(W)}{T}_{\mu}{}^{\sigma}{}_{\sigma}\delta_{\nu}{}^{\rho}$.

\subsection{Gauge redundancy and deformed Weitzenb\"ock connections}\label{GRgaugeredundancysection}

One might worry about the fact that the different connections we are obtaining are, in some sense, gauge artifacts of the Weitzenb\"ock connection.  We want to stress, again, that only when taking Lorentz transformations with constant parameters, the connection $\Gamma_{\mu\nu}{}^{\rho}$ and tensors built out of it are well defined, in the sense that they transform as a scalar under Lorentz transformations. And only when considering actions or equations of motion the local Lorentz symmetry is restored. This is completely analogous to the usual teleparallel gravity case. Thus, if we consider global Lorentz transformations it is clearly not possible to gauge-transform the different connections into the Weitzenb\"ock connection. Moreover, if we allowed for local  Lorentz transformations it is not possible neither. Take for instance the curvature tensor (\ref{RiemannFlat}) with the general connection (\ref{SpinConnectionTau}):

\begin{align}\nonumber
R_{a b c}{}^{d}(w)= &D_{a} \Big({a}_{1} {\tau}_{c b}\,^{d} - {b}_{1} {\tau}^{d}\,_{b c} - {b}_{2} {\tau}^{d}\,_{c b} + {c}_{1} {\eta}_{b c} {\tau}_{e}\,^{d e} \\\nonumber
&+ {d}_{1} {\delta}_{b}\,^{d} {\tau}_{c}\,^{e}\,_{e} + {d}_{2} {\delta}_{c}\,^{d} {\tau}_{b}\,^{e}\,_{e}\Big) \\\nonumber
&-D_{b} \Big({a}_{1} {\tau}_{c a}\,^{d} - {b}_{1} {\tau}^{d}\,_{a c} - {b}_{2} {\tau}^{d}\,_{c a}\\\nonumber
& + {c}_{1} {\eta}_{a c} {\tau}_{e}\,^{d e} + {d}_{1} {\delta}_{a}\,^{d} {\tau}_{c}\,^{e}\,_{e} + {d}_{2} {\delta}_{c}\,^{d} {\tau}_{a}\,^{e}\,_{e}\Big)\\\label{fullRiemmangaugeconn}
 &+ \cdots,
\end{align}   
where the dots represent quadratic terms. From the above derivative terms we see that the only possibility for the curvature tensor to vanish is that all of the coefficients must be set equal to zero, implying that the Weitzenb\"ock connection is the only connection written in terms of derivatives of the vielbein that makes the curvature tensor to vanish. Therefore, if the connections we have obtained so far were gauge-transformed (under \emph{local} Lorentz transformations) of the Weitzenb\"ock connection, it would imply a vanishing curvature tensor for these connections. As we have stated in the previous examples, this is not the case. More generally, different set of coefficients live in different Lorentz gauge orbits.   

On the other hand, we have mentioned about transformations that leave the curvature tensor or scalar curvature invariant, (\ref{gaugetransconncurvature}) and (\ref{gaugetransfRicciScalar}) respectively. From a geometric point of view, the connections are meant to be defined up to a ``gauge transformation'' of the form (\ref{gaugetransconncurvature}) as long as the new connection enjoys the same index symmetry properties as the old one. For instance, the Levi-Civita connection is symmetric in its two lower indices and if we apply the transformation (\ref{gaugetransconncurvature}) the symmetry in the two indices will be broken resulting in a new connection with torsion and non-metricity. The connections obtained in this work do not have a particular definite symmetry on their indices, and thus, transformations of the form (\ref{gaugetransconncurvature}) might be considered. However, none of our connections are related in this sense, since there is no combination of vielbein that produces a non-trivial scalar as in the last term of (\ref{gaugetransconncurvature}). The only transformation relevant for us is (\ref{gaugetransfRicciScalar}). In this case there are connections that are related as we have explicitly shown in (\ref{gammaweylspace}). Despite of this, the geometric properties defined by $\overset{(i)}{\widetilde{\Gamma}}$ are different from $\overset{(i)}{\Gamma}$: Clearly, $\overset{(i)}{\widetilde{\Gamma}}$ yields non-trivial non-metricity while $\overset{(i)}{\Gamma}$ does not. Also, the form of the curvature tensor for these connections are different since the $d_{2}$ term does not decouple from the curvature tensor (see (\ref{fullRiemmangaugeconn})).

Having clarified the gauge redundancy issue, the interpretation we are giving to the new connections is that they are deformed versions of the Weitzenb\"ock connection \cite{Hehl:1994ue}. Given two connections $\Gamma_{1}$ and $\Gamma_{2}$ we can always relate them by a tensor. Take $\Gamma_{1}=W + w_{1}$ and $\Gamma_{2}=W + w_{2}$ as in the decomposition (\ref{affineandgaugeconn}). Since $w$ is a tensor under diffeomorphisms, we obtain  $\Gamma_{1}=\Gamma_{2} + (w_{1} - w_{2})$. Thus, $\Gamma_{1}$ and $\Gamma_{2}$ are related by a tensor and we say $\Gamma_{1}$ is a deformed version of $\Gamma_{2}$. In this work, what we did is to find those $w$ parametrized by (\ref{SpinConnectionTau}) such that $\Gamma_{1}=W + w$ is a deformation of $\Gamma_{2}=W$ and reproduce the same action (or equations of motion) for the vielbein.

\subsection{Equations of Motion}\label{eqofmotionsection}

In this subsection we will obtain the equations of motion for the vielbein in terms of quantities associated to the connections we have obtained in the previous subsections. 

As a warm-up, we first obtain the equations of motion in the teleparallel case (i.e. $\overset{(4)}{w}_{a b}{}^{c}=0$) as it is usually done in the literature. However, we will apply a more general procedure in (\ref{subsecBGenericCaseEOM}) that will include the Weitzenb\"ock case as a particular case.

\subsubsection{Weitzenb\"ock case}\label{subunoEOMsec}

We are interested in the right hand side of (\ref{FundamentalEquationtwo}) using planar indices and the Weitzenb\"ock connection $\overset{(4)}{w}_{a b}{}^{c}=0$. For metric-compatible connections the non-metricity tensor vanishes, therefore the only contribution to the action comes from the contorsion tensor, i.e. $C_{a b}{}^{c}=K_{a b}{}^{c}$. Expanding in terms of the torsion we get:

\begin{align}\nonumber
&\overset{(4)}{K}_{a b}{}^{c}\overset{(4)}{K}_{c}{}^{a b} - \overset{(4)}{K}_{c b}{}^{c}\overset{(4)}{K}_{a}{}^{a b}=\\
&~~=-\frac{1}{4}\overset{(4)}{T}_{a b c} \overset{(4)}{T}{}^{a b c} -\frac{1}{2}\overset{(4)}{T}_{a b c} \overset{(4)}{T}{}^{a c b} + \overset{(4)}{T}_{a b}{}^{b} \overset{(4)}{T}{}^{a c}{}_{c}\,.
\end{align}
Where we recall that $\overset{(4)}{T}_{a b}{}^{c}=-\tau_{a b}{}^{c}$. Since the Riemann tensor associated to the Weitzenb\"ock connection vanishes, we expect to find equations of motion in terms of only $T_{a b}{}^{c}$. The right hand side of (\ref{FundamentalEquationtwo}) thus gives:
\begin{align}\nonumber
& \int dx^{4} e\Big(-\frac{1}{4}\overset{(4)}{T}_{a b c} \overset{(4)}{T}{}^{a b c} -\frac{1}{2}\overset{(4)}{T}_{a b c} \overset{(4)}{T}{}^{a c b} + \overset{(4)}{T}_{a b}{}^{b} \overset{(4)}{T}{}^{a c}{}_{c}\Big)=\\
&~~~= \int dx^{4} e\Big(-\frac{1}{2}\overset{(4)}{\widehat{T}}{}^{a b c} \overset{(4)}{T}_{a b c}\Big).
\end{align}
Here $e=\sqrt{-g}$ and in the last line we have introduced the tensor $\widehat{T}_{a b c}$ known as the superpotential in TG:
\begin{align}\nonumber
\hat{T}_{a b c}&=K_{c b a} + \eta_{a c} T_{b d}{}^{d} - \eta_{b c} T_{a d}{}^{d}\label{superpotential}\\
& = \frac{1}{2}\Big(T_{a b c} + 2 T_{[a |c| b]} + \eta_{a c} T_{b d}{}^{d} - \eta_{b c} T_{a d}{}^{d} \Big).
\end{align}
The superpotential satisfies $\widehat{T}_{a b c} = 2 \widehat{T}_{[a b] c}$. Due to this property the variation of the lagrangian is  easier to perform:
\begin{align}\nonumber
\delta_{e}\Big(-\frac{1}{2} e \overset{(4)}{\widehat{T}}{}^{a b c} \overset{(4)}{T}_{a b c} \Big) =&\frac{1}{2}e e_{d\mu}\delta e^{d\mu} \overset{(4)}{\widehat{T}}{}^{a b c} \overset{(4)}{T}_{a b c} \\
&~~~~ - e \overset{(4)}{\widehat{T}}{}^{a b c} \delta_e \overset{(4)}{T}_{a b c}\,,
\end{align}
where
\begin{align}\nonumber
- e \overset{(4)}{\widehat{T}}{}^{a b c} \delta_e \overset{(4)}{T}_{a b c} \, = & \,  e \overset{(4)}{\widehat{T}}{}^{a b c} \Big(2\delta e_{a}{}^{\mu}e_{d \mu} \Omega^{d}{}_{b c} + \tau_{a b}{}^{d} e_{d}{}^{\mu} \delta e_{c \mu}\\ \label{stvofaction}
& ~~~~~~~~~~~~~~~~ + 2 D_{a}\delta e_{b}{}^{\mu} e_{c \mu} \Big)\,.
\end{align}
Since the last term of (\ref{stvofaction}) is antisymmetrized on indices $(a,b)$ we can use:
\begin{equation}\label{DaCalDa}
2D_{[a}A_{b]}=2\accentset{(\{\})}{\mathfrak{D}}_{[a}A_{b]} + 2\overset{(\{\})}{w}_{[a b]}{}^{c}A_{c},
\end{equation}
where $\accentset{(\{\})}{\mathfrak{D}}_{a}A_{b}=D_{a}A_{b} - \overset{(\{\})}{w}_{a b}{}^{c} A_{c}$.  When substituting (\ref{DaCalDa}) in (\ref{stvofaction}) we see that the second term in the right-hand side of (\ref{DaCalDa}) cancels the $\tau$ term  in the first line of (\ref{stvofaction}) since $2\overset{(\{\})}{w}_{[a b]}{}^{c}=\tau_{a b}{}^{c}$ and $\delta e_{a\mu}e^{b\mu}=-e_{a\mu}\delta e^{b\mu}$. We proceed to use the product rule:
\begin{align}\nonumber
&e\overset{(4)}{\widehat{T}}{}^{a b c}2\accentset{(\{\})}{\mathfrak{D}}_{a}\delta e_{b}{}^{\mu} e_{c \mu} =\\\nonumber
&~~~=2e\accentset{(\{\})}{\mathfrak{D}}_{a}\Big(\overset{(4)}{\widehat{T}}{}^{a b c}\delta e_{b}{}^{\mu} e_{c \mu}\Big) - 2e\accentset{(\{\})}{\mathfrak{D}}_{c}\overset{(4)}{\widehat{T}}{}^{c a d}\delta e_{a}{}^{\mu} e_{d \mu}~~~ \\\label{ttofstvofaction}
&~~~~~~~ -2e\overset{(4)}{\widehat{T}}{}^{a b c}\delta e_{b}{}^{\mu} (-e_{e\mu} \Omega_{a}{}^{e}{}_{c} - \overset{(\{\})}{w}_{a c}{}^{e} e_{e \mu}).
\end{align}
The first term on the  right-hand side of (\ref{ttofstvofaction}) yields a total derivative and the last line of (\ref{ttofstvofaction}) comes  from expanding $\accentset{(\{\})}{\mathfrak{D}}_{a}e_{c\mu}$. Putting all together we then have:
\begin{align}\nonumber
&\delta_{e}\Big(-\frac{1}{2} e \overset{(4)}{\widehat{T}}{}^{a b c} \overset{(4)}{T}_{a b c} \Big) =\\\nonumber
&=\frac{1}{2}e e_{d\mu}\delta e^{d\mu} \overset{(4)}{\widehat{T}}{}^{a b c} \overset{(4)}{T}_{a b c}+ 2e \delta e_{a}{}^{\mu}e_{d \mu}\Big( \overset{(4)}{\widehat{T}}{}^{a b c} \tau^{d}{}_{b c}\\\label{variationLagW}
&~~~~~~~~~~~~~~~~~~~~~~~~~~~~ -\accentset{(\{\})}{\mathfrak{D}}_{c}\overset{(4)}{\widehat{T}}{}^{c a d}  + \overset{(4)}{\widehat{T}}{}^{b a c} \overset{(\{\})}{w}_{b c}{}^{d}\Big),
\end{align}
up to a total derivative. The $\Omega$ term of (\ref{ttofstvofaction}) combines with the $\Omega$ term of (\ref{stvofaction}) to form the $\tau$ term in the second line of (\ref{variationLagW}). The equations of motion are simply:
\begin{align}\nonumber
2 \accentset{(\{\})}{\mathfrak{D}}_{c} \overset{(4)}{\widehat{T}}{}^{c}{}_{a b}   + \,  2 \overset{(4)}{\widehat{T}}{}_{a c d} \Big( \overset{(4)}{T}{}_{b}{}^{c d} -  \overset{(4)}{K}{}^{c d}{}_{b} \Big) - \frac{1}{2} \eta_{a b} \overset{(4)}{T}_{c d e}\overset{(4)}{\widehat{T}}{}^{c d e} = 0.\\\label{eomTGflat}
\end{align}
Here we have used $\overset{(4)}{T}_{a b}{}^{c}=-\tau_{a b}{}^{c}$ and $\overset{(4)}{K}_{a b}{}^{c}=-\overset{(\{\})}{w}_{a b}{}^{c}$.
\subsubsection{Generic case}\label{subsecBGenericCaseEOM}
In the previous subsection we obtained the equations of motion by varying the right hand side of (\ref{FundamentalEquationtwo}). This is our goal here too. The variational procedure, however, of the preceding subsection depends heavily on the form of the torsion with respect to the vielbein and if we perform that procedure   for the new connections it is not easy to see how to rearrange terms in meaningful geometric quantities. In the Weitzenb\"ock case, the equations of motion are fully described in terms of the torsion since there is no curvature. In the case of the new connections,  all of them have non-zero curvature and we expect the Ricci tensor to appear in the equations of motion. Therefore, we will follow here a different route to derive the equations   of motion for our generic connections, and in particular, it will yield the equations of motion for the Weitzenb\"ock case (\ref{eomTGflat}). 

We start with the following action,
\begin{equation}\label{actiongenericcase}
S=\int e R(\Gamma) dx^4\,,
\end{equation}
where $R(\Gamma)=g^{\mu\nu} R_{\mu \rho \nu}{}^{\rho}(\Gamma)$. We are assuming   that $\Gamma$ is written in terms of first-order derivatives of the vielbein. We now vary the action with respect to the vielbein, 
\begin{align}\nonumber
&\delta S = \int \delta e R(\Gamma) dx^4\\\label{eqvarSgeneric}
 &~~~~~~~ + \int e \delta g^{\mu\nu} R_{\mu \rho \nu}{}^{\rho}(\Gamma) + \int e g^{\mu\nu} \delta  R_{\mu \rho \nu}{}^{\rho}(\Gamma).
\end{align}
The first term of (\ref{eqvarSgeneric}) is as usual,
\begin{equation}
-\int e e_{a \mu} R(\Gamma) \delta e^{a \mu}.
\end{equation}
The second term of (\ref{eqvarSgeneric}) gives:
\begin{equation}
2 \int e R_{(\mu |\rho| \nu)}{}^{\rho}(\Gamma) e^{a\nu} \delta e_{a}{}^{\mu}.
\end{equation}
The third term of (\ref{eqvarSgeneric}) can be decomposed as follows. We use the decomposition for $\Gamma$ as given by (\ref{eqDecompOfGamma}) implying a decomposition for the Riemann tensor   (\ref{eqDecompOfRiemann}). Thus, we obtain: 
\begin{equation}\label{eqvarSgenericthird}
\int e g^{\mu \nu} \delta \Big(R_{\mu \rho \nu}{}^{\rho}(\{\}) + 2\overset{(\{\})}{\nabla}_{[\mu}C_{\rho]\nu}{}^{\rho} + 2C_{[\mu |\lambda|}{}^{\rho} C_{\rho]\nu}{}^{\lambda}\Big)\,.
\end{equation}
The variation of the first term of (\ref{eqvarSgenericthird}) is   obtained by varying (\ref{RiemannCurved}) with respect to $\Gamma$ with $\Gamma$ being the Levi-Civita connetion. The result is:
\begin{equation}
\int e g^{\mu\nu} \overset{(\{\})}{\nabla}_{[\mu}\delta \overset{(\{\})}{\Gamma}_{\sigma]\nu}{}^{\sigma}=0.
\end{equation}
Which yields zero, since it is a total derivative. The second and third term of (\ref{eqvarSgenericthird}) can be rewritten as follows: 
\begin{align}\nonumber
&\int  e g^{\mu\nu}\,\delta\Big( 2\overset{(\{\})}{\nabla}_{[\mu}C_{\rho]\nu}{}^{\rho}\Big) = \\\nonumber
&= \delta \Big(\int  e g^{\mu\nu}\, ( 2\overset{(\{\})}{\nabla}_{[\mu}C_{\rho]\nu}{}^{\rho})\Big) -  \int \delta e g^{\mu\nu}\, ( 2\overset{(\{\})}{\nabla}_{[\mu}C_{\rho]\nu}{}^{\rho})\\
&~~~~~~ - \int e \delta g^{\mu\nu}\, ( 2\overset{(\{\})}{\nabla}_{[\mu}C_{\rho]\nu}{}^{\rho}), 
\end{align}
and
\begin{align}\nonumber
&\int  e g^{\mu\nu}\,\delta\Big( 2C_{[\mu |\lambda|}{}^{\rho} C_{\rho]\nu}{}^{\lambda}\Big) = \\\nonumber
&= \delta \Big(\int  e g^{\mu\nu}\, ( 2C_{[\mu |\lambda|}{}^{\rho} C_{\rho]\nu}{}^{\lambda})\Big) -  \int \delta e g^{\mu\nu}\, ( 2C_{[\mu |\lambda|}{}^{\rho} C_{\rho]\nu}{}^{\lambda})\\
&~~~~~~~~ - \int e \delta g^{\mu\nu}\, ( 2C_{[\mu |\lambda|}{}^{\rho} C_{\rho]\nu}{}^{\lambda}). 
\end{align}
Putting all together
\begin{align}\nonumber
\delta S = \int e \Bigg(g_{\mu\nu}\bigg( -R(\Gamma) + 2\overset{(\{\})}{\nabla}_{[\sigma} C_{\rho]}{}^{\sigma \rho} + 2C_{[\sigma|\lambda|}{}^{\rho} C_{\rho]}{}^{\sigma\lambda}\bigg)\, +   \\\nonumber
~~~~~+ \,2 R_{(\mu |\rho| \nu)}{}^{\rho}(\Gamma) - 2\overset{(\{\})}{\nabla}_{(\mu} C_{|\rho|\nu)}{}^{\rho} + 2 \overset{(\{\})}{\nabla}_{\rho} C_{(\mu\nu)}{}^{\rho}  \\\nonumber
~~~~ -\, 2C_{(\mu|\lambda}{}^{\rho} C_{\rho|\nu)}{}^{\lambda} + 2 C_{\rho\lambda}{}^{\rho} C_{(\mu\nu)}{}^{\lambda} \Bigg)e_{a}{}^{\nu} \delta e^{a\mu} \,+  \\\label{varactiongenericcase}
~~~~~~~~~~~~~~~~~~~~~~~+\, \delta \Bigg( 2\int e C_{[\mu |\lambda|}{}^{\rho} C_{\rho]}{}^{\mu\lambda} \Bigg)\,.
\end{align}
Since we are dealing   with $\Gamma$'s such that the Ricci scalar vanishes then the action (\ref{actiongenericcase}) vanishes. This implies that $\delta S =0$ off-shell. Therefore,   the right hand side of (\ref{varactiongenericcase}) must  vanish identically. On the other hand, the last term of (\ref{varactiongenericcase}) is equal to the variation of the right hand side of (\ref{FundamentalEquationtwo}) and we  need to impose this variation to vanish in order to obtain the equations of motion. So, the resulting (equivalent) equations of motion are:
\begin{align}\nonumber
&\,2 R_{(\mu |\rho| \nu)}{}^{\rho}(\Gamma)  + g_{\mu\nu}\Big(2\overset{(\{\})}{\nabla}_{[\sigma} C_{\rho]}{}^{\sigma \rho} + 2C_{[\sigma|\lambda|}{}^{\rho} C_{\rho]}{}^{\sigma\lambda}\Big)\\[1pt]\nonumber
&~~~ - 2\overset{(\{\})}{\nabla}_{(\mu} C_{|\rho|\nu)}{}^{\rho} + 2 \overset{(\{\})}{\nabla}_{\rho} C_{(\mu\nu)}{}^{\rho}  -\, 2C_{(\mu|\lambda}{}^{\rho} C_{\rho|\nu)}{}^{\lambda}  \label{eomgencasecurvedindices} \\[2pt]
&~~~ + 2 C_{\rho\lambda}{}^{\rho} C_{(\mu\nu)}{}^{\lambda}\, =\, 0, 
\end{align}
where we have used $R(\Gamma)=0$.
Equivalently, in flat indices the equations of motion are:
\begin{align}\nonumber
&\,2 R_{(a |c| b)}{}^{c}(w)  + \eta_{a b}\Big(2\accentset{(\{\})}{\mathfrak{D}}_{[d} C_{c]}{}^{d c} + 2C_{[d|e|}{}^{c} C_{c]}{}^{d e}\Big) \\[1pt]\nonumber
& ~~~ - 2\accentset{(\{\})}{\mathfrak{D}}_{(a} C_{|c|b)}{}^{c} + 2 \accentset{(\{\})}{\mathfrak{D}}_{c} C_{(a b)}{}^{c}  -\, 2C_{(a|e}{}^{c} C_{c|b)}{}^{e}  \\[2pt]\label{eomgenericvarflat}
&~~~ + 2 C_{c e}{}^{c} C_{(a b)}{}^{e}\, =\, 0. 
\end{align}
Equations (\ref{eomgencasecurvedindices}) and (\ref{eomgenericvarflat}) are the equations of motion for the general case (\ref{GammaConnectioncurvedsix}) and (\ref{SpinConnectionTau}) respectively. In the case of metric-compatible connections only the contorsion tensor contributes to $C_{a b}{}^{d}$. In this case, and after some manipulations, equation (\ref{eomgenericvarflat}) gets the form:
\begin{align}\nonumber
&\,2 R_{(a |c| b)}{}^{c}(w) + 2 \accentset{(\{\})}{\mathfrak{D}}_{c} \widehat{T}{}^{c}{}_{a b}   + \,  2 \widehat{T}{}_{a c d} \Big( T_{b}{}^{c d} -  K^{c d}{}_{b} \Big)  \\\nonumber
&~~~~~~~~~~~~~ - \frac{1}{2} \eta_{a b} T_{c d e}\widehat{T}{}^{c d e} + \accentset{(\{\})}{\mathfrak{D}}_{c} T_{a b}{}^{c} + 2\accentset{(\{\})}{\mathfrak{D}}_{[a} T_{b] c}{}^{c} \\\label{eomgenericvarflatTorsion}
&~~~~~~~~~~~~~ + T_{[b}{}^{c d} T_{|c d| a]} - T_{a b}{}^{c} T_{c d}{}^{d}\,=\,0, 
\end{align}
where $\widehat{T}_{a b}{}^{c}$ was defined in (\ref{superpotential}). It is easy to see that (\ref{eomgenericvarflatTorsion}) reduces to (\ref{eomTGflat}) for the Weitzenb\"ock case,
 $\overset{(4)}{w}_{a b}{}^{c}=0$, $R_{a b c}{}^{d}(\overset{(4)}{w})=0$, $\overset{(4)}{T}_{a b}{}^{c}= - \tau_{a b}{}^{c}$, with the help of the following identity:
\begin{equation}
\accentset{(\{\})}{\mathfrak{D}}_{c} \tau_{a b}{}^{c} + 2 \accentset{(\{\})}{\mathfrak{D}}_{[a} \tau_{b] c}{}^{c} -\tau_{[b}{}^{c d} \tau_{|c d| a]}  + \tau_{a b}{}^{c} \tau_{c d}{}^{d} = 0,
\end{equation}
As a final remark, we have verified that when plugging the connections (\ref{QzeroSolOneConn})-(\ref{QzeroSolFourConn}) inside equations (\ref{eomgenericvarflatTorsion}) reproduce the same equations of motion for the vielbein as in teleparallel gravity or Einstein gravity.

\section{Connections in Double Field Theory}\label{DFTsection2}
In this section we will try to apply the same procedure of the previous section (\ref{Notation}) to Double Field Theory and remark the similarities and differences. We briefly review the DFT ingredients we are going to use. Most of the details can be found in the literature mentioned in the Introduction.  

String theory has several dualities relating different theories defined on different backgrounds or regimes of validity. In particular, T-duality is a duality relating string theory compactified on different toroidal backgrounds. The T-duality symmetry is represented by the $O(n,n)$ group, where $n$ represents the number of internal compactified dimensions. Double Field Theory is a proposal to incorporate the T-duality group as a manifest symmetry of a field theory. It achieves this by doubling the usual space-time coordinates, incorporating new coordinates $\tilde{x}$ dual to the winding of the string. The fields of DFT are represented by a generalized metric $H_{MN}(X)$ and generalized dilaton $d(X)$, where $X^M=(\tilde{x},x)$ are the DFT space-time coordinates, $M=1,...,2D$ (where  $D$ is the target space-time dimension containing both external and internal coordinates). Indices are raised and lowered with the invariant $O(D,D)$ metric,

\begin{equation}\label{eqDFTetacurved}
\eta_{MN} = \begin{pmatrix} 0 & 1\\ 1 & 0\end{pmatrix}.
\end{equation}
      
The generalized metric satisfies $H_{M}{}^{N} H_{N P} = \eta_{MP}$ and $H_{MN}=H_{NM}$. DFT is a constrained theory, it usually satisfies the strong constraint $\partial_{M}\partial^M (...)=0$, where the dots represents arbitrary products  of fields and gauge parameters. However, there are instances where it is known the strong constraint can be relaxed, for example in generalized Scherk-Schwarz compactifications (see, for instance, \cite{Aldazabal:2013sca} and references therein). For simplicity, we will assume the strong constraint in this work. It is convenient to introduce projectors $P$ and $\bar{P}$:
\begin{align}\label{dftprojone}
P_{M}{}^N = \frac{1}{2}\big( \delta_M{}^N - H_M{}^N \big),&~~~\bar{P}_{M}{}^N = \frac{1}{2}\big( \delta_{M}{}^N + H_M{}^N\big),\\\label{dftprojtwo}
P_{M}{}^M + \bar{P}_{M}{}^N = \delta_{M}{}^N,&~~~P_{M}{}^P\bar{P}_{PN}=0.
\end{align}

These projectors allow us to separate components of tensors in orthogonal subspaces and we will use the notation $V_{\dbar{M}}=P_{M}{}^NV_N$ and $V_{\bar{M}}=\bar{P}_{M}{}^NV_N$. It is possible to introduce a generalized vielbein $E_A{}^M$ as in GR: 

\begin{equation}
H_{MN} = S_{AB} E^{A}{}_M E^{B}{}_N,~~~\eta_{MN} = \eta_{AB}E^{A}{}_ME^{B}{}_N,
\end{equation}
where
\begin{equation}
S_{AB} = \begin{pmatrix} s^{ab} & 0\\ 0 & s_{ab}\end{pmatrix},~~~a,b=1,...,D,~~~s_{ab}=diag(-,+,...,+),
\end{equation}
and $\eta_{AB}$ is numerically equivalent to (\ref{eqDFTetacurved}). The generalized vielbein satisfies $E_{A}{}^M E_{BM} = \eta_{AB}$, $E_{AM}E^{A}{}_{N}=\eta_{MN}$. Here, we have introduced planar indices $A,B,C,D,E,F=1,...,2D$ which are raised and lowered with $\eta_{AB}$. The curved indices ranges from $K,L,M,N,...=1,...,2D$. The theory is invariant under generalized coordinate transformations acting on curved indices:   
\begin{align}\nonumber
\delta_\xi E^{A}{}_M & =\mathcal{L}_\xi E^A{}_M\\
&=\xi^P\partial_PE^A{}_M + (\partial_M\xi^P - \partial^P\xi_M)E^A{}_P,
\end{align}
and a local double Lorentz transformation  $O(1,D-1)\times O(1,D-1)$ acting on the planar ones:
\begin{align}
\delta_\Lambda E_A{}^M & = \Lambda_A{}^B E_{B}{}^M. 
\end{align}
Here $\xi^M=\xi^M(X)$ is an infinitesimal parameter of coordinate transformations and $\Lambda_{A}{}^B=\Lambda_A{}^B(X)$ satisfies $\Lambda_{AB} = - \Lambda_{BA}$, $S_A{}^{A'}\Lambda_{A'B} = S_{B}{}^{B'}\Lambda_{AB'}$ (following the conventions of \cite{Geissbuhler:2013uka}). The dilaton transforms under generalized coordinate transformations as $\delta_{\xi}d = \xi^P\partial_Pd - \frac{1}{2}\partial_P\xi^P$.
Of course, the theory is also invariant under global $O(D,D)$ transformations acting on  curved indices. 

It is possible in DFT to introduce concepts like connections and curvatures by introducing a covariant derivative under generalized coordinate transformations:
\begin{equation}
\nabla_{M}V_N = \partial_MV_N -\Gamma_{MN}{}^PV_P.
\end{equation}

The DFT connection transforms in the following way under generalized coordinate transformations:
\begin{equation}
\delta_{\xi}\Gamma_{MN}{}^P=\mathcal{L}_{\xi}\Gamma_{MN}{}^P + \partial_{M}\partial_{N}\xi^P - \partial_{M}\partial^P\xi_N.
\end{equation}
The concepts of covariant derivative and generalized coordinate transformation seems to parallel the one in standard geometry reviewed in section (\ref{Notation}). In particular, we saw that a generic connection in a metric-affine space-time can be decomposed in terms of the appropriate geometric quantities as in (\ref{eqDecompOfGamma}). This decomposition has allowed us to study and find, in a systematic way, different connections and equations of motion based on different geometric quantities. In the following subsections we would like to find a similar decomposition for a generic generalized DFT connection that helps us to unravel the different geometric quantities allowed by the theory. That is, we would like to find a decomposition of the form:

\begin{equation}\label{eqDecompOfGammaDFT}
\Gamma_{M N}{}^{P} = \mbox{\scriptsize{$\left\{\begin{array}{@{}c@{}} P \\ M ~N\end{array}\right\}$}} + K_{M N}{}^{P} + L_{M N}{}^{P},
\end{equation}
where $\mbox{\scriptsize{$\left\{\begin{array}{@{}c@{}} P \\ M ~N\end{array}\right\}$}}$ would play the role of generalized Levi-Civita connection and the rest would be the generalization of the contorsion and L-tensor.  Let us introduce two quantities that transform as tensors under generalized coordinate transformations which can be interpreted as the generalization of the torsion tensor in general relativity:
\begin{equation}\label{eqtau1}
\mathcal{T}^{(1)}_{MP}{}^Q \equiv \Gamma_{MP}{}^Q - \Gamma_{PM}{}^Q + \eta^{QR}\eta_{TP}\Gamma_{RM}{}^T\,,
\end{equation}
\begin{align}\nonumber
\mathcal{T}^{(2)}_{MP}{}^Q & \equiv \Gamma_{MP}{}^Q +  \eta^{QR}\eta_{TP}\Gamma_{MR}{}^T\\\label{eqtau2}
& \equiv - \eta^{QR}\nabla_{M}\eta_{PR}\,.
\end{align}
The tensor $\mathcal{T}^{(1)}$ in (\ref{eqtau1}) has been named the generalized torsion tensor in DFT. The tensor $\mathcal{T}^{(2)}$  in (\ref{eqtau2}) is the representation of the covariant derivative of $\eta_{MN}$ which a priori we are not setting to zero. Now we have all the tools to find a decomposition as in (\ref{eqDecompOfGammaDFT}).

\subsection{Tensors $Q$ and \texorpdfstring{$\bar{Q}$}{} }\label{DFTsection2QandbarQ}
Since there are two kinds of metrics in DFT we need to study the covariant derivative of both of them. Nevertheless, instead of working with them, we will work with projectors (\ref{dftprojone}):

\begin{equation}\label{eqnablaP}
\nabla_{M}P_{NP}=\partial_{M}P_{NP}-\Gamma_{MN\dbar{P}} -\Gamma_{MP\dbar{N}},
\end{equation}

\begin{equation}\label{eqnablabarP}
\nabla_{M}\bar{P}_{NP}=\partial_{M}\bar{P}_{NP}-\Gamma_{MN\bar{P}} -\Gamma_{MP\bar{N}}.
\end{equation}

For simplicity, it will be convenient to introduce the following quantities:
\begin{equation}
Q_{MNP}=\nabla_{M}P_{NP},~~~\bar{Q}_{MNP}=\nabla_M\bar{P}_{NP}.
\end{equation}
They play the role of the  non-metricity tensor in Riemannian geometry. From equation (\ref{eqnablaP}) we can hit with $\bar{P}$ on the $N$ index to obtain:

\begin{equation}\label{gammaconstr1}
\Gamma_{M\bar{N}\dbar{P}}=\bar{P}_N{}^{N'}\partial_M P_{N'P} - Q_{M\bar{N}P}.
\end{equation}
And similarly with (\ref{eqnablabarP}):
\begin{equation}\label{gammaconstr2}
\Gamma_{M\dbar{N}\bar{P}}=P_N{}^{N'}\partial_M \bar{P}_{N'P} - \bar{Q}_{M\dbar{N}P}.
\end{equation}
Equations (\ref{gammaconstr1}) and (\ref{gammaconstr2}) fully determine $\Gamma_{\dbar{M}\bar{N}\dbar{P}}$, $\Gamma_{\bar{M}\bar{N}\dbar{P}}$ and $\Gamma_{\dbar{M}\dbar{N}\bar{P}}$  $\Gamma_{\bar{M}\dbar{N}\bar{P}}$, respectively.    

Note also that the following components are equal to zero:

\begin{equation}
\bar{Q}_{M\dbar{P}\dbar{Q}}=P_P{}^{P'}P_{Q}{}^{Q'}\nabla_{M}\bar{P}_{P'Q'}=0,
\end{equation}
\begin{equation}
Q_{M\bar{P}\bar{Q}}=\bar{P}_P{}^{P'}\bar{P}_{Q}{}^{Q'}\nabla_{M}P_{P'Q'}=0.
\end{equation}
This can be seen by using the orthogonality of the projectors and the Leibniz rule.

\subsection{Components from $\mathcal{T}^{(1)}$}\label{DFTsection2gentorsion}
We would like to point out something that has not been stated in the literature so far. If we take $\mathcal{T}^{(1)}_{MN}{}^P=0$ then:
\begin{equation}
\Gamma^Q{}_{MP}=-\left(\Gamma_{MP}{}^Q-\Gamma_{PM}{}^Q\right),~~\Rightarrow \Gamma_{QMP}=\Gamma_{Q[MP]}.
\end{equation}
Thus,  with the above imposition $\Gamma$ turns out to be antisymmetric in its last two indices and then:
\begin{equation}
\mathcal{T}^{(2)}_{MNP}=-\nabla_{M}\eta_{NP}=2\Gamma_{M(NP)}=0.
\end{equation} 
So, imposing $\mathcal{T}^{(1)}=0$ already implies the constraint $\nabla\eta=0$. This shows that both $\mathcal{T}^{(1)}$ and $\mathcal{T}^{(2)}$ are related. 

Returning to the undetermined components, from the definition of $\mathcal{T}^{(1)}$ we can obtain new determined components for the connection by projecting the indices:
\begin{equation}\label{eqfoura}
\Gamma_{\dbar{P}\bar{M}\bar{Q}}=-\mathcal{T}^{(1)}_{\bar{M}\dbar{P}\bar{Q}} + \Gamma_{\bar{M}\dbar{P}\bar{Q}} + \Gamma_{\bar{Q}\bar{M}\dbar{P}},
\end{equation}  
\begin{equation}\label{eqfourb}
\Gamma_{\bar{M}\dbar{P}\dbar{Q}}=\mathcal{T}^{(1)}_{\bar{M}\dbar{P}\dbar{Q}} + \Gamma_{\dbar{P}\bar{M}\dbar{Q}} - \Gamma_{\dbar{Q}\bar{M}\dbar{P}}
\end{equation} 
The above components for the connection are fully determined in terms of the physical fields and $\mathcal{T}^{(1)}$.

From the tensor $\mathcal{T}^{(2)}$ we obtain no new information since we note that can be written in several ways using previous tensors:
\begin{equation}\label{eqtau2intermsof1}
\begin{split}
\mathcal{T}^{(2)}_{MPQ}&=2\Gamma_{M(PQ)}\\
& = 2\mathcal{T}^{(1)}_{(PQ)M}\\
&=-Q_{MPQ}-\bar{Q}_{MPQ}
\end{split}
\end{equation}
This shows that $\mathcal{T}^{(2)}$ is not independent from $\mathcal{T}^{(1)}$ or the $Q$'s tensors. And, of course, also shows, that this particular combination of $\mathcal{T}^{(1)}$ is related to the $Q$'s.
Moreover, not all of the components of $\mathcal{T}^{(1)}$ are independent. By using the definition of $\mathcal{T}^{(1)}$ and  (\ref{eqfoura}) and (\ref{eqfourb}), we can see which components of $\mathcal{T}^{(1)}$ are related . The relations are as follows:

\begin{equation}
\mathcal{T}^{(1)}_{\dbar{M}\bar{N}\dbar{P}} = -\mathcal{T}^{(1)}_{\bar{N}\dbar{M}\dbar{P}} - Q_{\dbar{P}\bar{N}\dbar{M}} - \bar{Q}_{\dbar{P}\dbar{M}\bar{N}}
\end{equation}

\begin{equation}
\mathcal{T}^{(1)}_{\dbar{M}\bar{N}\bar{P}} = -\mathcal{T}^{(1)}_{\bar{N}\dbar{M}\bar{P}} - Q_{\bar{P}\bar{N}\dbar{M}} - \bar{Q}_{\bar{P}\dbar{M}\bar{N}}
\end{equation}

\begin{equation}\nonumber\begin{split}
\mathcal{T}^{(1)}_{\dbar{M}\dbar{N}\bar{P}} = & \mathcal{T}^{(1)}_{\bar{P}\dbar{M}\dbar{N}} - \bar{Q}_{\dbar{M}\dbar{N}\bar{P}} - 
Q_{\dbar{M}\bar{P}\dbar{N}}\\
&+\bar{Q}_{\dbar{N}\dbar{M}\bar{P}} + Q_{\dbar{N}\bar{P}\dbar{M}}
\end{split}
\end{equation}

\begin{equation}
\mathcal{T}^{(1)}_{\bar{M}\bar{N}\dbar{P}} = -\mathcal{T}^{(1)}_{\bar{M}\dbar{P}\bar{N}} - Q_{\bar{M}\bar{N}\dbar{P}} - \bar{Q}_{\bar{M}\dbar{P}\bar{N}}
\end{equation}

We choose the components $\mathcal{T}^{(1)}_{\bar{M}\dbar{N}\bar{P}}$ and $\mathcal{T}^{(1)}_{\bar{M}\dbar{N}\dbar{P}}$  as independent ones and built the others from them.

\subsection{Components from $\nabla e^{-2d}$}\label{DFTsection2nabladilaton}

We define the covariant derivative acting on $e^{-2d}$ as
\begin{equation}\label{eqnabladilaton}
\nabla_{M}e^{-2d} = \partial_Me^{-2d} - \Gamma_{QM}{}^Qe^{-2d}.
\end{equation}
From this equation we can determine the trace components of the connection, since $\Gamma_{QM}{}^Q=\Gamma_{\dbar{Q}M}{}^{\dbar{Q}} + \Gamma_{\bar{Q}M}{}^{\bar{Q}}$. So by projecting the free index we obtain:

\begin{equation}\label{tracegammadbar}
\Gamma_{\dbar{Q}\dbar{M}}{}^{\dbar{Q}} = -\Gamma_{\bar{Q}\dbar{M}}{}^{\bar{Q}} + e^{2d}P_M{}^{M'}\left(-\nabla_{M'}e^{-2d} + \partial_{M'}e^{-2d}\right)
\end{equation}
\begin{equation}\label{tracegammabar}
\Gamma_{\bar{Q}\bar{M}}{}^{\bar{Q}} = -\Gamma_{\dbar{Q}\bar{M}}{}^{\dbar{Q}} + e^{2d}\bar{P}_M{}^{M'}\left(-\nabla_{M'}e^{-2d} + \partial_{M'}e^{-2d}\right)
\end{equation}
The trace components are then fully determined. These are the only equations from where we may try to obtain information about $\Gamma_{\dbar{M}\dbar{N}\dbar{P}}$ and $\Gamma_{\bar{M}\bar{N}\bar{P}}$, the equations from previous sections do not yield any non-trivial information about them. 
Following \cite{Hohm:2011si} we propose the following ansatz for $\Gamma_{\dbar{M}\dbar{N}\dbar{P}}$ and $\Gamma_{\bar{M}\bar{N}\bar{P}}$:
\begin{align}\nonumber
\Gamma_{\dbar{M}\dbar{N}\dbar{P}} & = - \frac{2}{1-D}P_{M[N}P_{P]}{}^{P'} \Gamma_{\bar{Q}\dbar{P'}}{}^{\bar{Q}} + \tilde{\Gamma}_{\dbar{M}\dbar{N}\dbar{P}} + \Sigma_{\dbar{M}\dbar{N}\dbar{P}}~~~+\\\label{gamma3down}
&~~~~+ \frac{2e^{2d}}{1-D}P_{M[N}P_{P]}{}^{P'}\big( -\nabla_{P'}e^{-2d} + \partial_{P'}e^{-2d}  \big).
\end{align}
\begin{align}\nonumber
\Gamma_{\bar{M}\bar{N}\bar{P}} & = - \frac{2}{1-D}\bar{P}_{M[N}\bar{P}_{P]}{}^{P'}\Gamma_{\dbar{Q}\bar{P'}}{}^{\dbar{Q}} + \tilde{\Gamma}_{\bar{M}\bar{N}\bar{P}} +\Sigma_{\bar{M}\bar{N}\bar{P}} ~~~+\\\label{gamma3up}
&~~~~+ \frac{2e^{2d}}{1-D}\bar{P}_{M[N}\bar{P}_{P]}{}^{P'}\big( -\nabla_{P'}e^{-2d} + \partial_{P'}e^{-2d}  \big).
\end{align}
Note that the trace parts of the connection on the right hand sides  are determined components given by the trace of (\ref{gammaconstr1}) and (\ref{gammaconstr2}). The ansatz (\ref{gamma3down}) and (\ref{gamma3up}) satisfy  (\ref{tracegammadbar}) and (\ref{tracegammabar}) provided we introduce the quantities  $\tilde{\Gamma}$ and $\Sigma$ that satisfy:
\begin{equation}\label{dftconditgammatilde1}
\tilde{\Gamma}_{\dbar{Q}\dbar{M}}{}^{\dbar{Q}} = 0,~~\tilde{\Gamma}_{\bar{Q}\bar{M}}{}^{\bar{Q}} = 0,
\end{equation}
\begin{equation}\label{dftconditSigmatensor1}
\Sigma_{\dbar{Q}\dbar{M}}{}^{\dbar{Q}} = 0,~~\Sigma_{\bar{Q}\bar{M}}{}^{\bar{Q}} = 0.
\end{equation}
And also:
\begin{equation}\begin{split}\label{dftconditgammatilde2}
0 & = \tilde{\Gamma}_{\dbar{Q}\dbar{M}\dbar{R}} - \tilde{\Gamma}_{\dbar{M}\dbar{Q}\dbar{R}} + \tilde{\Gamma}_{\dbar{R}\dbar{Q}\dbar{M}},\\
\end{split}
\end{equation}
\begin{equation}\label{dftconditSigmatensor2}\begin{split}
\mathcal{T}^{(1)}_{\dbar{Q}\dbar{M}\dbar{R}} & = \Sigma_{\dbar{Q}\dbar{M}\dbar{R}} - \Sigma_{\dbar{M}\dbar{Q}\dbar{R}} + \Sigma_{\dbar{R}\dbar{Q}\dbar{M}},
\end{split}
\end{equation}
And similarly for $\tilde{\Gamma}_{\bar{M}\bar{N}\bar{P}}$ and $\Sigma_{\bar{M}\bar{N}\bar{P}}$. 
The above requirements come from (\ref{eqtau1}) and (\ref{eqtau2intermsof1}). The quantity $\tilde{\Gamma}$ was introduced in \cite{Hohm:2011si} and it is an undetermined component in terms of the physical fields ($P_{MN}, \bar{P}_{MN},d$). It is responsible for fixing the transformation properties of $\Gamma_{\dbar{M}\dbar{N}\dbar{P}}$ or  $\Gamma_{\bar{M}\bar{N}\bar{P}}$ under generalized coordinate transformations. Here, we need to introduce a new quantity, $\Sigma$, which is a tensor under generalized coordinate transformations. Depending on the choice for the tensor $\mathcal{T}^{(1)}$ (which might carry physical degrees of freedom), the tensor $\Sigma$ may be fully expressed in terms of it. This means, to express $\Sigma$ as a linear combination of $\mathcal{T}^{(1)}$ (we will see how to do this for particular cases in subsections (\ref{DFTsection2TDFT}) and  (\ref{DFTsection2exampotherconn})).

\subsection{The full connection}\label{DFTsection2fullconnection}
  By gathering together  (\ref{gammaconstr1}),(\ref{gammaconstr2}), (\ref{eqfoura}),  (\ref{eqfourb}), (\ref{gamma3down}) and (\ref{gamma3up}) we are now able to write the full connection. The final expression is: 

\begin{equation}\label{DFTfullconnectionLCContL}
\Gamma_{M N P} =  \mbox{\scriptsize{$\left\{\begin{array}{@{}c@{}} P \\ M ~N\end{array}\right\}$}} + K_{M N P} + L_{M N P},
\end{equation}
with
\begin{align}\nonumber
 \mbox{\scriptsize{$\left\{\begin{array}{@{}c@{}} P \\ M ~N\end{array}\right\}$}} = & -\frac{2}{1-D}P_{M[N}P_{P]}{}^{P'}\left(\bar{P}^{M'N'}P_{P'}{}^R\partial_{M'}\bar{P}_{N'R}  - e^{2d}\partial_{P'}e^{-2d}\right)\\\nonumber
&  -\frac{2}{1-D}\bar{P}_{M[N}\bar{P}_{P]}{}^{P'}\left(P^{M'N'}\bar{P}_{P'}{}^R\partial_{M'}P_{N'R}  - e^{2d}\partial_{P'}e^{-2d}\right)\\\nonumber
& + P_{M}{}^{M'}P_{N}{}^{N'}\partial_{M'}\bar{P}_{N'P} + P_{M}{}^{M'}\bar{P}_{N}{}^{N'}\partial_{M'}P_{N'P}\\\nonumber
& + \bar{P}_{N}{}^{N'}P_{M}{}^{M'}\partial_{N'}\bar{P}_{M'P} + \bar{P}_{P}{}^{P'}\bar{P}_{N}{}^{N'}\partial_{P'}P_{N'M} \\\nonumber
& + \bar{P}_{M}{}^{M'}\bar{P}_{N}{}^{N'}\partial_{M'}P_{N'P} + \bar{P}_{M}{}^{M'}P_{N}{}^{N'}\partial_{M'}\bar{P}_{N'P}\\\nonumber
& + P_{N}{}^{N'}\bar{P}_{M}{}^{M'}\partial_{N'}P_{M'P} - P_{P}{}^{P'}\bar{P}_{M}{}^{M'}\partial_{P'}P_{M'N} \\\label{dftgenlevcivconn}
& + \tilde{\Gamma}_{\dbar{M}\dbar{N}\dbar{P}}+ \tilde{\Gamma}_{\bar{M}\bar{N}\bar{P}},
\end{align}
\begin{align}\label{DFTconttensor}
K_{MNP} = -\mathcal{T}^{(1)}_{\bar{N}\dbar{M}\bar{P}} + \mathcal{T}^{(1)}_{\bar{M}\dbar{N}\dbar{P}} + \Sigma_{\dbar{M}\dbar{N}\dbar{P}} + \Sigma_{\bar{M}\bar{N}\bar{P}}
\end{align}
\begin{align}\nonumber
L_{MNP} = & -\frac{2e^{2d}}{1-D}P_{M[N}P_{P]}{}^{P'}\nabla_{P'}e^{-2d}-\frac{2e^{2d}}{1-D}\bar{P}_{M[N}\bar{P}_{P]}{}^{P'}\nabla_{P'}e^{-2d}\\\label{DFTLtensor}
& + \frac{2}{1-D}P_{M[N}P_{P]}{}^{P'}\bar{Q}_{\bar{Q}\dbar{P'}}{}^{\bar{Q}} + \frac{2}{1-D}\bar{P}_{M[N}\bar{P}_{P]}{}^{P'}Q_{\dbar{Q}\bar{P'}}{}^{\dbar{Q}}\\\nonumber
& -\bar{Q}_{\dbar{M}\dbar{N}\bar{P}} - Q_{\dbar{M}\bar{N}\dbar{P}} - \bar{Q}_{\bar{N}\dbar{M}\bar{P}} - Q_{\bar{P}\bar{N}\dbar{M}}\\\nonumber
& - Q_{\bar{M}\bar{N}\dbar{P}} - \bar{Q}_{\bar{M}\dbar{N}\bar{P}} - Q_{\dbar{N}\bar{M}\dbar{P}} + Q_{\dbar{P}\bar{M}\dbar{N}}.
\end{align}
We have obtained the generalized Levi-Civita connection $\mbox{\scriptsize{$\left\{\begin{array}{@{}c@{}} P \\ M ~N\end{array}\right\}$}}$, which after some manipulations is equal to the connection obtained in \cite{Hohm:2011si}.  The generalized Levi-Civita connection has vanishing torsion, it is $\eta$-$H$-compatible, i.e. $\nabla P=\nabla\bar{P}=0$ and has vanishing $\nabla e^{-2d}=0$. We have also defined the generalized  contorsion tensor (\ref{DFTconttensor}), and the $L$-tensor (\ref{DFTLtensor}) made up of the non-metricity tensors $Q$, $\bar{Q}$ and $\nabla e^{-2d}$.  
For simplicity, we define 
\begin{equation}
C_{MNP}\equiv K_{MNP} + L_{MNP}.
\end{equation}
Note that it is the tensor $C_{MNP}$ the one that accounts for equation (\ref{eqtau1}) rather than only using the generalized contorsion tensor $K_{MNP}$, that is
\begin{equation}
\mathcal{T}^{(1)}_{MNP} = C_{MNP} - C_{NMP} + C_{PMN}.
\end{equation}
In a metric-affine geometry only the contorsion tensor is needed to yield the torsion, i.e. $T_{\mu\nu}{}^{\rho}=2\Gamma_{[\mu\nu]}{}^{\rho}=2K_{[\mu\nu]}{}^{\rho}$. 
\subsection{Generalized Riemann tensor}\label{DFTsection2genriemanntensor}

The results shown here are a little bit more general compared to the ones stated before in the literature since we do not assume any extra index property on the Riemann tensor (a priori) given by imposition of constrains in the connection. The generalized Riemann tensor in curved index is given by:

\begin{equation}
\mathcal{R}_{MNKL} = R_{MNKL} + R_{KLMN} + \Gamma_{QMN}\Gamma^{Q}{}_{KL},
\end{equation}
with
\begin{equation}\label{dftRofgenriemanntensor}
R_{MNKL} = \eta_{QL}\big(2\partial_{[M}\Gamma_{N]K}{}^Q + 2\Gamma_{[M|R|}{}^Q\Gamma_{N]K}{}^R \big).
\end{equation}

By definition, the generalized Riemann tensor satisfies $\mathcal{R}_{MNKL} = \mathcal{R}_{KLMN}$. The generalized Riemann tensor can be written in planar indices by rotating with the generalized vielbein and the below quantity:
\begin{equation}
\Omega_{ABC} = E_{A}{}^M\partial_ME_{B}{}^N E_{CN}
\end{equation}
Note that $\Omega_{ABC} = - \Omega_{ACB}$. Of course, we can use the generalized vielbein $E_{A}{}^M$ to convert curved indices to planar ones.  We will often use $\Omega_{MNP}=E^A{}_{M}E^{B}{}_{N}E^{C}{}_{P}\Omega_{ABC}$. We define the ``generalized fluxes":
\begin{equation}
\mathcal{F}_{ABC} \equiv 3 \Omega_{[ABC]},
\end{equation}
\begin{equation}
\mathcal{F}_{A} \equiv \Omega^B{}_{BA} + 2D_{A}d,
\end{equation}
where $D_A = E_{A}{}^M\partial_M$. It is important to stress that the generalized fluxes (with planar indices) transform as scalars under generalized coordinate transformations when imposing the strong constraint. The relation between the connection and the gauge connection is:

\begin{equation}\label{eqdftrelGammaWandw}
\Gamma_{MN}{}^P = W_{MN}{}^P + E^A{}_NE_{B}{}^Pw_{MA}{}^B.
\end{equation} 
where $W_{MN}{}^P = -\Omega_{MN}{}^P$ is the generalized Weitzenb\"ock connection and we will refer to $w_{ABC}$ as the DFT gauge connection. This gauge connection transforms as a scalar (with all planar indices) under general coordinate transformations. In principle, one can require that $w_{MBC}$ transforms as a gauge connection under local double Lorentz transformation. However, as stated in the Introduction, we are only concerned with  its transformation properties under generalized coordinate transformations since the local double Lorentz group will be generically broken, i.e., only global double Lorentz transformations (that is with constant parameters) will survive at the connection level. The generalized curvature tensor in planar indices then gets the form:

\begin{equation}\label{dftgenriemanncurvplanar}
\mathcal{R}_{ABCD} = \hat{R}_{ABCD} + \hat{R}_{CDAB} + w^E{}_{AB} w_{ECD},
\end{equation} 
where 
\begin{equation}
\hat{R}_{ABCD} = 2D_{[A}w_{B]CD} - \mathcal{F}_{AB}{}^E w_{ECD} - 2 w_{[A|C|}{}^E w_{B]ED}\,.
\end{equation}

Under a variation of the connection the Riemann tensor changes to\footnote{Since a priori we are not assuming $\nabla \eta = 0$ we will work with a connection defined with its third index up $\Gamma_{MN}{}^K$ just like in Riemannian geometry. However, we will raise and lower its  indices with $\eta$ when this poses no problem.}:

\begin{equation}\label{DFTGammavar}
\Gamma'_{MN}{}^K = \Gamma_{MN}{}^K + \delta \Gamma_{MN}{}^K,
\end{equation}
\begin{equation}\label{DFTRiemannvar}\begin{split}
\Rightarrow ~~\delta\mathcal{R}_{MNKL}  = &  2\eta_{LQ}\overset{(\Gamma)}{\nabla}_{[M}\delta\Gamma_{N]K}{}^Q + \overset{(\Gamma)}{\mathcal{T}^{(1)}}_{MN}{}^Q\delta\Gamma_{QKL}\\
& + 2\eta_{NQ}\overset{(\Gamma)}{\nabla}_{[K}\delta\Gamma_{L]M}{}^Q +  \overset{(\Gamma)}{\mathcal{T}^{(1)}}_{KL}{}^Q\delta\Gamma_{QMN}.\\
\end{split}
\end{equation}

We now know that we can decompose a general connection in terms of the generalized Levi-Civita connection, the generalized contorsion and the L-tensor:
\begin{equation}
\Gamma_{M N}{}^{P} =  \mbox{\scriptsize{$\left\{\begin{array}{@{}c@{}} P \\ M ~N\end{array}\right\}$}} + C_{M N}{}^{P},
\end{equation}

When plugging the above decomposition in the Riemann tensor we obtain the general form: 

\begin{equation}\begin{split}
\mathcal{R}_{MNKL}(\Gamma)  = & \mathcal{R}_{MNKL}(\{\}) + 2\eta_{LQ}\overset{(\{\})}{\nabla}_{[M}C_{N]K}{}^Q + \overset{(\{\})}{\mathcal{T}^{(1)}}_{MN}{}^QC_{QKL}\\
& + 2\eta_{NQ}\overset{(\{\})}{\nabla}_{[K}C_{L]M}{}^Q +  \overset{(\{\})}{\mathcal{T}^{(1)}}_{KL}{}^QC_{QMN}\\
&+ 2C_{[M|QL|}C_{N]K}{}^Q + 2C_{[K|QN|}C_{L]M}{}^Q\\\label{dftgencurvatensorgeneric}
& + C_{QMN}C^{Q}{}_{KL}. 
\end{split}
\end{equation}

Where we have used $\mathcal{R}_{MNKL}=\mathcal{R}_{KLMN}$. When the connection is equal to the generalized Levi-Civita, it is possible from the generalized Riemann curvature to define a generalized Ricci tensor $\mathcal{R}_{\dbar{M}\dbar{N}\bar{K}}{}^{\dbar{N}}$ and  a Ricci scalar $\mathcal{R}=P^{MK}P^{NL}\mathcal{R}_{MNKL}$. The equations of motion of DFT are just the vanishing of the generalized Ricci tensor and the vanishing of the generalized Ricci scalar, which gives the equations of motion for the generalized metric $H_{MN}$ and the dilaton $d$, respectively. These equations can be found by varying the DFT action:

\begin{align}\nonumber
S &= \int dX e^{-2d}\mathcal{R}(\{\})\\\nonumber
& = \int dX e^{-2d}\Big(4H^{MN}\partial_{M}\partial_{N}d - \partial_{M}\partial_{N}H^{MN} + 4\partial_{M}H^{MN}\partial_{N}d - 4H^{MN}\partial_{M}d\partial_Nd\\\label{DFTactionintermsofHandd}
&~~~~~~~~~~~~~~~~~~~~ -\frac{1}{2}H^{MN}\partial_{M}H^{KL}\partial_{K}H_{NL} + \frac{1}{8}H^{MN}\partial_{M}H^{KL}\partial_{N}H_{KL}\Big).
\end{align}

A couple of Bianchi identities can be deduced from the generalized Riemann tensor \cite{Hohm:2011si,Geissbuhler:2013uka}. All of them can be expressed in terms of generalized contorsion, L-tensor and non-vanishing curvature for generic connections using (\ref{dftgencurvatensorgeneric}). For instance, a B.I for full antisymmetrization on the generalized Riemann tensor can be translated as:

\begin{align}\nonumber
0= \mathcal{R}_{[MNKL]}(\{\})= &   \, \mathcal{R}_{[MNKL]}(\Gamma) - 4\eta_{Q[L}\overset{(\{\})}{\nabla}_{M}C_{NK]}{}^Q\\
&~~~~- 4C_{[M|Q|L}C_{NK]}{}^Q - C_{Q[MN}C^{Q}{}_{KL]}. 
\end{align}
Where $\Gamma$ is a generic generalized connection and we have used that $\overset{(\{\})}{\mathcal{T}^{(1)}}=0$. In the flux formulation, the above B.I is important since it can be related to a B.I for the fluxes
\begin{align}\nonumber
\mathcal{R}_{[ABCD]}(\{\})=\frac{4}{3}D_{[A}\mathcal{F}_{BCD]} - \mathcal{F}_{[AB}{}^{E}\mathcal{F}_{CD]E}=0.
\end{align} 

Before closing this subsection we would like to make a comment on extracting information about components of the generalized connection using a possible Palatini formulation. It is known that in GR, one can use the Palatini formalism to obtain equations of motion for the connection and for the metric separately, and when one solves for the connection and plugs the solution into the equation of motion of the metric, the Einstein's equation of motion is recovered. It is not our intention to apply a Palatini formalism to DFT but just to vary the DFT action with respect to the connection and see what information we get from there. We consider the action

\begin{equation}\label{DFTgenericscalarcurvature}
S = \int dX e^{-2d} P^{MK}P^{NL}\mathcal{R}_{MNKL}(\Gamma).
\end{equation}
Then we vary with respect to $\Gamma$ using (\ref{DFTGammavar})- (\ref{DFTRiemannvar}). We then have

\begin{equation}\begin{split}
e^{-2d}P^{MK}P^{NL}\delta \mathcal{R}_{MNKL} & = e^{-2d}P^{[M|K|}P^{N]}{}_L\Big(4 \nabla_M\delta \Gamma_{NK}{}^L\Big)\\
& ~~~ +\, 2e^{-2d}P^{MK}P^{NL}\mathcal{T}^{(1)}_{MN}{}^Q\delta\Gamma_{QKL}\,.
\end{split}
\end{equation}  
We then plug the above calculation in (\ref{DFTgenericscalarcurvature}) and after an integration by parts and recalling the definition of $\nabla e^{-2d}$ we obtain:
\begin{equation}
S=\int dX\Big[-4\nabla_M\Big( e^{-2d}P^{[M|K|}P^{N]}{}_L\Big) + 2 e^{-2d}P^{MK}P^{Q}{}_L\mathcal{T}^{(1)}_{MQ}{}^N \Big]\delta\Gamma_{NK}{}^L\,.
\end{equation}

We thus obtain an equation relating the non-metricity and the generalized torsion:

\begin{equation}\label{eqDFTvargammaintoaction}
0=-4\nabla_M\Big(e^{-2d}P^{[M|K|}P^{N]}{}_L\Big) + 2 e^{-2d}P^{MK}P^{Q}{}_L\mathcal{T}^{(1)}_{MQ}{}^N\,.
\end{equation}

This is something to expect. Just to compare, if we performed the same calculation in GR we would obtain: 
\begin{align}\nonumber\label{grvargamma}
0 & =T_{\tau\lambda}{}^\mu + \nabla_{\lambda}g^{\mu\epsilon}g_{\tau\epsilon} - g^{\epsilon\theta}\nabla_{[\tau}g_{|\epsilon\theta|}\delta^{\mu}_{\lambda]}\\
& ~~~-g_{\epsilon\tau}\nabla_{\sigma}g^{\sigma\epsilon}\delta^{\mu}_{\lambda} - 2T_{[\tau|\rho}{}^{\rho}\delta^{\mu}_{\lambda]}\,.
\end{align}
In both cases, the torsion and non-metricity get related. In GR, when the connection is metric compatible the solution to (\ref{grvargamma}) is given by $T_{\mu\nu}{}^{\rho}=0$. In the DFT case, if $P=\frac{1}{2}(\eta - H)$ in (\ref{eqDFTvargammaintoaction}) and the connection is $\eta$-$H$-compatible then only some components of the generalized torsion are required to be equal to zero but not all of them, a priori. We have analyzed several components of (\ref{eqDFTvargammaintoaction}) and they seem to give no new information compared to the analysis of the previous subsection.

\subsection{Equations of Motion}\label{DFTsection2EoM}
We start from the general action

\begin{equation}
S = \int dX e^{-2d} P^{MK}P^{NL}\mathcal{R}_{MNKL}(\Gamma).
\end{equation}
To find the equations of motion we vary with respect to the vielbein $E_{A}{}^M$ and the dilaton. With respect to the vielbein:

\begin{align}
\delta_{E} S & = \int dX e^{-2d}\delta_E \mathcal{R}(\Gamma)\\
& = 2\int dX e^{-2d}\delta_E P^{MK} P^{NL} \mathcal{R}_{MNKL}(\Gamma) + \int dX e^{-2d}P^{MK}P^{NL}\delta_E\mathcal{R}_{MNKL}(\Gamma),
\end{align}
and\footnote{The reader is referred to \cite{Geissbuhler:2013uka} for this result. The important point is that it only makes use of $\mathcal{R}_{MNKL}=\mathcal{R}_{KLMN}$.}
\begin{equation}
2(\delta_E P^{MK})P^{NL}\mathcal{R}_{MNKL}(\Gamma) = -4\Delta_{AC}E^A{}_{M} E^{C}{}_{N} P^{QN}\bar{P}^{RM}\mathcal{R}_{Q\dbar{P}R}{}^{\dbar{P}}
\end{equation}

where

\begin{equation}
\Delta_{AB} = \delta E_A{}^M E_{BM} = -\Delta_{BA}.
\end{equation}

The other term of the action

\begin{align}\nonumber
\int e^{-2d} &  P^{MK} P^{NL}  \delta_E \mathcal{R}_{MNKL}(\Gamma)=\\\nonumber
& = \int e^{-2d}P^{MK}P^{NL}\Bigg[ \delta_E \mathcal{R}_{MNKL}(\{\})\\\nonumber
&~~~~ + \delta_E \Big( 2\eta_{LQ}\overset{(\{\})}{\nabla}_{[M}C_{N]K}{}^Q + 2\eta_{NQ}\overset{(\{\})}{\nabla}_{[K}C_{L]M}{}^Q \Big)\\\label{eqvar2dotermGRDFT}
& ~~~~+  \delta_E \Big( C^2_{MNKL}\Big)  \Bigg]
\end{align}

Where we have defined for simplicity:

\begin{equation}
C^2_{MNKL} = 2C_{[M|QL|}C_{N]K}{}^Q + 2C_{[K|QN|}C_{L]M}{}^Q
+ C_{QMN}C^{Q}{}_{KL} 
\end{equation}

 Now, the first term on the right hand side of (\ref{eqvar2dotermGRDFT}) can be shown to be a total derivative just like in GR (see \cite{Geissbuhler:2013uka,Hohm:2011si} for more details) so we focus on the other two terms. They can be cast in the following form, respectively:

\begin{equation}\nonumber
 \int e^{-2d} P^{MK}P^{NL} \delta_E\Big( 2\eta_{LQ}   \overset{(\{\})}{\nabla}_{[M}C_{N]K}{}^Q + 2\eta_{NQ}\overset{(\{\})}{\nabla}_{[K}C_{L]M}{}^Q \Big)=~~~~~~~~~
\end{equation}
\begin{equation}\label{eqvar2dotermGRDFT1}
= -2\int e^{-2d}(\delta_E P^{MK}) P^{NL}  \Big( 2\eta_{LQ}\overset{(\{\})}{\nabla}_{[M}C_{N]K}{}^Q\\
 + 2\eta_{NQ}\overset{(\{\})}{\nabla}_{[K}C_{L]M}{}^Q \Big) 
\end{equation}  

and

\begin{align}\nonumber
\int e^{-2d}P^{MK}P^{NL}\delta_E C^2_{MNKL} 
& =\delta_E \int e^{-2d}P^{MK}P^{NL}C^2_{MNKL}\\
&~~~~~~~~-2\int e^{-2d}(\delta_EP^{MK})P^{NL}C^2_{MNKL}
\end{align}

%
%

In several steps we have made use of 
\begin{equation}
	\overset{(\{\})}{\mathcal{T}^{(1)}}{}_{MNP}=0,
\end{equation}
\begin{equation}
	\overset{(\{\})}{\nabla}_M P_{NQ} = \overset{(\{\})}{\nabla}_M \bar{P}_{NQ} = 0,
\end{equation}
\begin{equation}
	\overset{(\{\})}{\nabla}_{M}e^{-2d}= 0.
\end{equation}
These equations also allowed us to discard  a total derivative in (\ref{eqvar2dotermGRDFT1}).

Putting all together we have

\begin{align}\nonumber
\delta_E S & = -4\int e^{-2d}\Delta_{AC}E^A{}_ME^C{}_NP^{QN}\bar{P}^{RM}\Big(\mathcal{R}_{Q\dbar{P}R}{}^{\dbar{P}}(\Gamma) ~~~~~~ \\\nonumber
& ~~~~~~ - 2\overset{(\{\})}{\nabla}_{(Q}C_{|\dbar{P}|R)}{}^{\dbar{P}} + 2\overset{(\{\})}{\nabla}_{P}C_{(QR)}{}^{\dbar{P}} - C^2_{Q\dbar{P}R}{}^{\dbar{P}}\Big)\\\label{varRDFT}
& ~~~~~ + \delta_E \Big(\int e^{-2d}P^{MK}P^{NL}C^2_{MNKL} \Big)
\end{align}
The argument to find the equations of motion follows exactly as in subsection (\ref{eqofmotionsection}). We know that $\delta_E S=0$ off-shell since $\mathcal{R}(\Gamma)=0$ so the right hand side of (\ref{varRDFT}) vanishes identically. On the other hand, and due to  $\mathcal{R}(\Gamma)=0$ again, we know that up to a boundary term:

\begin{equation}\label{identDFTwithC2}
\int dX e^{-2d}\mathcal{R}(\{\}) = - \int dX e^{-2d}P^{MK}P^{NL}C^2_{MNKL}.
\end{equation} 

When varying with respect to $E_A{}^{M}$ we find the equation of motion and we see that the right hand side of (\ref{identDFTwithC2}) matches the last term of (\ref{varRDFT}). Therefore, equation of motion for $E_A{}^M$ from (\ref{varRDFT}) is:

\begin{align}\label{mastereqDFTforGamma}
0=P^{Q[N}\bar{P}^{M]R}\Big(\mathcal{R}_{Q\dbar{P}R}{}^{\dbar{P}}(\Gamma)
 - 2\overset{(\{\})}{\nabla}_{(Q}C_{|\dbar{P}|R)}{}^{\dbar{P}} + 2\overset{(\{\})}{\nabla}_{P}C_{(QR)}{}^{\dbar{P}} - C^2_{Q\dbar{P}R}{}^{\dbar{P}}\Big).
\end{align}

This is the equation of motion for the generalized vielbein in terms of the geometric quantities furnished by a generic generalized connection that renders null the generalized scalar curvature. We will see below the particular case of the generalized Weitzenb\"ock connection. Note also the similarity between (\ref{mastereqDFTforGamma}) and equation (\ref{eomgencasecurvedindices}) in the the GR case. The second term  on the first line of (\ref{eomgencasecurvedindices}) is actually absent in (\ref{mastereqDFTforGamma}) because the role of the determinant of the metric in GR is now played by $e^{-2d}$ in DFT and it is an independent degree of freedom. 

The procedure to find the dilaton equations of motion proceeds very closely as before, so we just quote the result:


\begin{align}\label{mastereqdftdilaton}
0= P^{MK} P^{NL}\Big(2\overset{(\{\})}{\nabla}_{[M}C_{N]KL} + 2\overset{(\{\})}{\nabla}_{[K}C_{L]MN} + C^2_{MNKL}\Big).
\end{align}

\subsection{Determination of undetermined parts of the Connection}\label{DFTsection2undetpartconn}

Before giving examples of connections that furnish equal theories to the standard DFT we would like to make an observation about the undetermined parts of the generalized Levi-Civita connection (\ref{dftgenlevcivconn}). Let us concentrate on particular projected components that contain undetermined parts:

\begin{align}\nonumber
\overset{(\{\})}{\Gamma}_{\dbar{M}\dbar{N}\dbar{P}} & = P_{M}{}^{M'} P_{N}{}^{N'} P_{P}{}^{P'}   \mbox{\scriptsize{$\left\{\begin{array}{@{}c@{}} P' \\ M' ~N'\end{array}\right\}$}} \\\nonumber
& = -\frac{2}{1-D}P_{M[N}P_{P]}{}^{P'}\big(\bar{P}^{M'N'}P_{P'}{}^{R}\partial_{M'}\bar{P}_{N'R} - e^{2d}\partial_{P'}e^{-2d}\big)\\\label{eq3dbarGLCcurv}
& ~~~~ + \tilde{\Gamma}_{\dbar{M}\dbar{N}\dbar{P}}\,.
\end{align} 

This is the only projected part of the full connection (together with $\Gamma_{\bar{M}\bar{N}\bar{P}}$) that possesses undetermined parts\footnote{For the generalized Levi-Civita case we can just set the undetermined part $\Sigma$ to zero, since the generalized torsion vanishes in this case.}, i.e. $\tilde{\Gamma}$. We recall that we have obtained this expression by demanding that 

\begin{equation}\label{condglvc}
\nabla_M P_{NP} = \nabla_M \bar{P}_{NP} = \mathcal{T}^{(1)}_{MNP} = \nabla_M e^{-2d} = 0.
\end{equation}

We know that a general connection can be rewritten in terms of the Weitzenb\"ock connection and a gauge connection $w_{MA}{}^B$ through (\ref{eqdftrelGammaWandw}), and when doing so, we can translate the conditions (\ref{condglvc}) as conditions on the gauge connection:

\begin{equation}
w_{ABC} + w_{ACB} = 0,~~S_{C}{}^{C'}w_{ABC'} + S_{B}{}^{B'}w_{ACB'} = 0,
\end{equation}
\begin{equation}
3w_{[ABC]}=\mathcal{F}_{ABC},~~w^{B}{}_{BA}=\mathcal{F}_A.
\end{equation}

Let us now try to solve these equations assuming a connection written in terms of first order derivatives of the generalized vielbein and dilaton such that $w_{ABC}$ is a scalar. This forces to write $w_{ABC}$ in terms of linear combinations of fluxes. The unique solution to the above equations is given by:
\begin{align}\nonumber
\overset{(\{\})}{w}_{ABC} & = \frac{1}{1-D}(\eta_{A[C}\mathcal{F}_{B]} + S_{A[C}S^D{}_{B]}\mathcal{F}_{D})\\\nonumber
& ~~~~+ \frac{1}{3}\Big(\mathcal{F}_{ABC} + S_{B}{}^{B'}S_{C}{}^{C'}\mathcal{F}_{AB'C'} \\\label{dftspinconnec}
& ~~~~ - \frac{1}{2}S_{A}{}^{A'}S_{B}{}^{B'}\mathcal{F}_{A'B'C} - \frac{1}{2}S_{A}{}^{A'}S_{C}{}^{C'}\mathcal{F}_{A'B C'}  \Big)
\end{align}

This spin connection does not transform well under local double Lorentz transformations but under constant ones. It is straightforward to check that this connection yields the usual generalized Ricci scalar in terms of the fluxes, that is
\begin{align}\nonumber
\mathcal{R}(\overset{(\{\})}{w}) & =S^{AB}(2D_{A}\mathcal{F}_{B}-\mathcal{F}_{A}\mathcal{F}_{B}) + \mathcal{F}_{ABC}\mathcal{F}_{DEF}\left(\frac{1}{4}S^{AD}\eta^{BE}\eta^{CF}-\frac{1}{12}S^{AD}S^{BE}S^{CF}\right)\\\label{eqDFTwithspinDFT}
& ~~~~ - 2D^A\mathcal{F}_{A} + \mathcal{F}^{A}\mathcal{F}_A - \frac{1}{6}\mathcal{F}^{ABC}\mathcal{F}_{ABC}.
\end{align}
The last line of (\ref{eqDFTwithspinDFT}) actually vanishes since we are imposing the strong constraint. The resulting action giving by $\mathcal{R}(\overset{(\{\})}{w})$ coincides with (\ref{DFTactionintermsofHandd}) up to a boundary term.
Now, we would like to point out that a seeming paradox comes about. Since $w_{ABC}$ defined by (\ref{dftspinconnec}) is a generalized scalar we can construct a fully determined connection by using (\ref{eqdftrelGammaWandw}). Let us call this connection $\Gamma'$: 

\begin{equation}\label{glcvprime}
\Gamma'_{MNP} = W_{MNP} +  \overset{(\{\})}{w}_{MNP},
\end{equation}
 where $\overset{(\{\})}{w}_{MNP} = E_{A}{}^{M}E_{B}{}^N  E_{C}{}^P \overset{(\{\})}{w}_{ABC}$. Let us see the implication of comparing (\ref{eq3dbarGLCcurv}) to  (\ref{glcvprime}) with all indices projected:

\begin{equation}\label{compglcctoprime}
\overset{(\{\})}{\Gamma}_{\dbar{M}\dbar{N}\dbar{P}} = \Gamma'_{\dbar{M}\dbar{N}\dbar{P}}.
\end{equation}

When comparing both sides of (\ref{compglcctoprime}) we see that the first line of (\ref{dftspinconnec}) gets canceled by several terms of the first line in (\ref{eq3dbarGLCcurv}) and the difference requires:

\begin{align}\label{detgammatilde}
\tilde{\Gamma}_{\dbar{M}\dbar{N}\dbar{P}} & = -\frac{2}{1-D}P_{M[N}P_{P]}{}^R\Omega^{\dbar{Q}}{}_{\dbar{Q}R}  -\Omega_{\dbar{M}\dbar{N}\dbar{P}} + \frac{1}{3}\mathcal{F}_{\dbar{M}\dbar{N}\dbar{P}}.
\end{align}
It is straight-forward to check that $\tilde{\Gamma}$ satisfies the desired requisites $\tilde{\Gamma}_{\dbar{M}\dbar{N}}{}^{\dbar{M}}=0$ and $\tilde{\Gamma}_{\dbar{M}\dbar{N}\dbar{P}} -\tilde{\Gamma}_{\dbar{N}\dbar{M}\dbar{P}} + \tilde{\Gamma}_{\dbar{P}\dbar{M}\dbar{N}}=0$ established in equations (\ref{dftconditgammatilde1}, \ref{dftconditgammatilde2}. The undetermined part $\tilde{\Gamma}_{\bar{M}\bar{N}\bar{P}}$ gets also fully determined with a similar form: 

\begin{equation}\label{detgammatildeup}
\tilde{\Gamma}_{\bar{M}\bar{N}\bar{P}}=-\frac{2}{1-D}\bar{P}_{M[N}\bar{P}_{P]}{}^{P'}\Omega^{\bar{Q}}{}_{\bar{Q}P'} - \Omega_{\bar{M}\bar{N}\bar{P}} + \frac{1}{3}\mathcal{F}_{\bar{M}\bar{N}\bar{P}}.
\end{equation}

The undetermined component $\tilde{\Gamma}$ of the generalized Levi-Civita connection is now determined in terms of physical fields (i.e. $E_{A}{}^M$ and $d$). The reason why it was no possible to fully determine the connection before is that demanding the connection to be only written in terms of $P_{MN}$ or $\bar{P}_{MN}$  is too restrictive. Indeed, the right-hand side of (\ref{detgammatilde}) (or (\ref{detgammatildeup})) cannot be fully written in terms of $P_{MN}$ (or $\bar{P}_{MN}$) and derivatives acting on them only. One way to see this is by noting that the first term of (\ref{detgammatilde}) should come from a combination of terms of the form $P_{NP}\partial_RP^{RQ}P_{QM}$, $P_{MP}\partial_RP^{RQ}P_{QN}$ and $P_{MN}\partial_RP^{RQ}P_{QP}$ (other combinations give zero). However, it is not possible with combinations of these terms to obtain exactly (\ref{detgammatilde}).

We would like at this point to make a small summary of the above situation and a comparison with general relativity.  We know that the GR's Levi-Civita connection is fully expressed in terms of first-order derivatives acting on the metric. The spin connection (\ref{SpinConnLevCivTau}), on the other hand, is written in terms of first-order derivatives acting on the vielbein, and through (\ref{affineandgaugeconn}), all the terms rearrange in such a way that only the metric appears. This can be done because the Weitzenb\"ock connection  cannot be written in terms of first-order derivatives acting on the metric. Besides, the GR's spin connection  transforms like a $SO(1,3)$-connection under local Lorentz transformations. We also know that the equations of motion of GR can be written either in terms of the metric $g_{\mu\nu}$ or the vielbein $e_{a}{}^{\mu}$ and when written in terms of the latter, a local Lorentz invariance is manifest. Therefore, the equations of motion of GR written in both languages contain exactly the same information. 

In DFT, the situation is very similar but, still, there are some differences. As we have seen above, the generalized Levi-Civita connection cannot be fully written in terms of $P_{MN}$ (or $\bar{P}_{MP}$) and that causes an undetermined part of the connection to be added when this choice is made. But if we make the choice to write the connection in terms of first-order derivatives acting on the generalized vielbein $E_{A}{}^M$ rather than $P_{MN}$ (or $\bar{P}_{MP}$), then, the undetermined parts get determined (\ref{detgammatilde})-(\ref{detgammatildeup}). From the curved DFT space-time point of view, the connection (\ref{glcvprime}) is well-defined (i.e. transforms as a connection under generalized coordinate transformations), it is fully determined and is unique. From the double Lorentz point of view only transformations with constant parameters are consistent. Note that in TG a similar situation occurs with the Weitzenb\"ock connection, where it is well-defined under diffeomorphisms but ill-behaved under local Lorentz transformations. 

Despite the fact that the generalized Levi-Civita connection (\ref{glcvprime}) can be fully written in terms of physical fields only if one chooses the generalized vielbein and dilaton as the fundamental fields, the DFT equations of motion can be fully written either in terms of $(P_{MN}, \bar{P}_{MN},d)$ or $(E_{A}{}^M,d)$. In the later case the local double Lorentz group is restored and the dynamical degrees of freedom are exactly the same in both cases\footnote{This is because the components $\tilde{\Gamma}_{\dbar{M}\dbar{N}\dbar{P}}$ and $\tilde{\Gamma}_{\bar{M}\bar{N}\bar{P}}$ are projected out from the generalized Ricci scalar, playing no role in the EoMs.}. 
In this sense, we note that this global-local mechanism is the same type of situation we have mentioned in the Introduction and analyzed throughout this article: theories that have a well-defined curved connection with respect to diffeomorphisms and with respect to constant Lorentz transformations, and yet, giving rise to equations of motion with a local gauge symmetry (i.e. parameters now depending on the coordinates). Therefore, the generalized Levi-Civita might also be interpreted as a deformed version of the generalized Weitzenb\"ock connection.    

One more comment before finishing this subsection. In DFT (and also in TG) the gauge connection $w_{ABC}$ is not really a connection in the classical sense. However, from the geometric space-time viewpoint this poses no problem since diffeomorphisms are not affected by this construction. In the TG case, the Weitzenb\"ock connection can be interpreted as the parallelization of the connection (given by the basis of vielbeins). In the  DFT case a proper geometric interpretation is lacking (see however some comments in the next section). 
\subsection{Teleparallel Double Field Theory}\label{DFTsection2TDFT}

With the machinery developed above it is now straightforward to consider particular cases for connections that renders different geometric theories compared to the usual DFT (which uses the generalized Levi-Civita connection). Of course, these geometric theories will describe the same dynamics for the generalized vielbein and generalized dilaton as the usual DFT. An important particular connection is the Weitzenb\"ock connection, which will be obtained as a particular solution in the next section. This connection is equal to 

\begin{equation}\label{dftweitconn}
\overset{(W)}{\Gamma}_{MN}{}^P = W_{MN}{}^P,~~~W_{MN}{}^P = -\Omega_{MN}{}^P.
\end{equation}

This connection is $\eta$-$H$-compatible, or equivalently 
\begin{equation}
\overset{(W)}{\nabla}_M P_{NP} = 0,~~~\overset{(W)}{\nabla}_M \bar{P}_{NP} = 0.
\end{equation}
 It can be interpreted as the analog of being metric-compatible in GR. The Weitzenb\"ock connection also renders the curvature tensor null, i.e. $\mathcal{R}_{MNPQ}(W)=0$. These properties would indicate that there is a distant parallelism in the DFT-space-time, or in other words, parallel transport of  vectors would not depend on the path. Indeed, it can be checked:

\begin{equation}
\overset{(W)}{\nabla}_M E_A{}^{N} = 0 \,.
\end{equation}

It is easy to see that the Weitzenb\"ock connection furnishes non-trivial generalized torsion and non-trivial $\nabla e^{-2d}$:

\begin{equation}\label{dftweitgtorsion}
\overset{(W)}{\mathcal{T}^{(1)}}_{MNP}=-\mathcal{F}_{MNP}
\end{equation}
\begin{equation}
\overset{(W)}{\nabla}_M e^{-2d} = - \mathcal{F}_M e^{-2d}
\end{equation}

Therefore, there is a non-trivial generalized contorsion tensor $K_{MNP}$ and also a non-trivial trace\footnote{One may wonder if it is possible to find a  connection such that $\mathcal{R}(\Gamma)=0$ and that also satisfies $\nabla P = \nabla \bar{P}= \nabla e^{-2d} = 0$ but with non-trivial generalized torsion. We found the answer to be no.} of $L_{MNP}$ due to $\nabla e^{-2d}\neq 0$. The fact that it is not enough with assuming only a non-trivial contorsion tensor contribution is a new feature compared to the usual Weitzenb\"ock connection of standard geometry. The connection (\ref{dftweitconn}) is fully determined in terms of the physical fields $(E_{A}{}^B,d)$:

\begin{align}\label{dftweitgcontorsion}
\overset{(W)}{K}_{MNP} = -\overset{(W)}{\mathcal{T}^{(1)}}_{\bar{N}\dbar{M}\bar{P}} + \overset{(W)}{\mathcal{T}^{(1)}}_{\bar{M}\dbar{N}\dbar{P}} + \overset{(W)}{\Sigma}_{\dbar{M}\dbar{N}\dbar{P}} + \overset{(W)}{\Sigma}_{\bar{M}\bar{N}\bar{P}}
\end{align}
\begin{equation}\begin{split}
\overset{(W)}{L}_{MNP} = & -\frac{2e^{2d}}{1-D}P_{M[N}P_{P]}{}^{P'}\overset{(W)}{\nabla}_{P'}e^{-2d}-\frac{2e^{2d}}{1-D}\bar{P}_{M[N}\bar{P}_{P]}{}^{P'}\overset{(W)}{\nabla}_{P'}e^{-2d}
\end{split}
\end{equation}

\begin{equation}\label{detsigmadft}
\overset{(W)}{\Sigma}_{\dbar{M}\dbar{N}\dbar{P}}=\frac{1}{3}\overset{(W)}{\mathcal{T}^{(1)}}_{\dbar{M}\dbar{N}\dbar{P}},~~\overset{(W)}{\Sigma}_{\bar{M}\bar{N}\bar{P}}=\frac{1}{3}\overset{(W)}{\mathcal{T}^{(1)}}_{\bar{M}\bar{N}\bar{P}}
\end{equation}

Equation (\ref{detsigmadft}) shows that, indeed, the generalized contorsion tensor (\ref{dftweitgcontorsion}) is fully described in terms of the generalized torsion tensor (\ref{dftweitgtorsion}). Thus, the theory and the geometry will be described through   $\overset{(W)}{\mathcal{T}^{(1)}}_{MNP}$ and $\overset{(W)}{\nabla}_M e^{-2d}$ rather than the generalized curvature of DFT. More explicitly, with the machinery developed in the last section, it is straightforward to write the equations of motion in terms of geometric quantities (like the generalized contorsion tensor which contains the generalized torsion and the tensor $L_{MNP}$ which contains $\nabla e^{-2d}$) given by the generalized Weitzenb\"ock connection. From (\ref{mastereqDFTforGamma}) and (\ref{mastereqdftdilaton}) the equations of motion for the Teleparallel equivalent of DFT are:

\begin{align}\label{dftTGeqE}
0 & =P^{Q[N}\bar{P}^{M]R}\Big(- 2\overset{(\{\})}{\nabla}_{(Q}\overset{(W)}{C}_{|\dbar{P}|R)}{}^{\dbar{P}} + 2\overset{(\{\})}{\nabla}_{P}\overset{(W)}{C}_{(QR)}{}^{\dbar{P}} - \overset{(W)}{C^2}_{Q\dbar{P}R}{}^{\dbar{P}}\Big),\\
0 & = P^{MK} P^{NL}\Big(2\overset{(\{\})}{\nabla}_{[M}\overset{(W)}{C}_{N]KL} + 2\overset{(\{\})}{\nabla}_{[K}\overset{(W)}{C}_{L]MN} + \overset{(W)}{C^2}_{MNKL}\Big),
\end{align}
where
\begin{align}\nonumber
\overset{(W)}{C}_{MNP} & = \overset{(W)}{K}_{MNP} + \overset{(W)}{L}_{MNP}\\\nonumber 
& =  -\overset{(W)}{\mathcal{T}^{(1)}}_{\bar{N}\dbar{M}\bar{P}} + \overset{(W)}{\mathcal{T}^{(1)}}_{\bar{M}\dbar{N}\dbar{P}} + \frac{1}{3}\overset{(W)}{\mathcal{T}^{(1)}}_{\dbar{M}\dbar{N}\dbar{P}} + \frac{1}{3}\overset{(W)}{\mathcal{T}^{(1)}}_{\bar{M}\bar{N}\bar{P}}\\
&~~~~  -\frac{2e^{2d}}{1-D}P_{M[N}P_{P]}{}^{P'}\overset{(W)}{\nabla}_{P'}e^{-2d}-\frac{2e^{2d}}{1-D}\bar{P}_{M[N}\bar{P}_{P]}{}^{P'}\overset{(W)}{\nabla}_{P'}e^{-2d}.
\end{align}

We would like to stress that when writing (\ref{dftTGeqE}) in flat indices and re-expressing it explicitly in terms of $\mathcal{T}^{(1)}$ and $\nabla_M e^{-2d}$, and after some algebra, the equation of motion gets a very similar form compared to the equation of motion of TG (\ref{eomTGflat}). That is, equation (\ref{dftTGeqE}) gets the form\footnote{It may be useful to use the following identity 
	\begin{equation}
	D^{C}\overset{(W)}{\mathcal{T}^{(1)}}_{CAB} + 2D_{[A}(e^{2d}\overset{(W)}{\nabla}_{B]} e^{-2d}) + (e^{2d}\overset{(W)}{\nabla}_C e^{-2d})\overset{(W)}{\mathcal{T}^{(1)C}}{}_{AB}= 0.
	\end{equation} }:

\begin{equation}\label{eqTEDFTintorsionterms}
0 = -2 S^{D[A}D^{B]}(e^{2d}\overset{(W)}{\nabla}_De^{-2d}) - (e^{2d}\overset{(W)}{\nabla}_De^{-2d} - D_D)\hat{\mathcal{T}}^{D[AB]} + \hat{\mathcal{T}}^{CD[A}\overset{(W)}{\mathcal{T}^{(1)}}_{CD}{}^{B]}
\end{equation}

Where $\overset{(W)}{\nabla}_D = E_D{}^{M}\overset{(W)}{\nabla}_{M}$. We have introduced the notation

\begin{equation}\label{DFTsuperpotential}
\hat{\mathcal{T}}^{ABC} = S^{ABCDEF}\overset{(W)}{\mathcal{T}^{(1)}}_{DEF}
\end{equation}

where 

\begin{equation}
S^{ABCDEF} = \frac{1}{2}S^{AD}\eta^{BE}\eta^{CF} + \frac{1}{2}\eta^{AD}S^{BE}\eta^{CF} + \frac{1}{2}\eta^{AD}\eta^{BE}S^{CF} - \frac{1}{2}S^{AD}S^{BE}S^{CF}.
\end{equation}

Here, $\hat{\mathcal{T}}_{ABC}=\hat{\mathcal{T}}_{[ABC]}$ because $\overset{(W)}{\mathcal{T}^{(1)}}$ is fully antisymmetric in its indices. We remark again that the generalized torsion transforms only under constant double Lorentz transformations but (\ref{eqTEDFTintorsionterms}) (which is the same as the DFT action in the vielbein formalism) possesses a local double Lorentz symmetry. The tensor 6-index tensor $S$ satisfies $S_{ABCDEF}S^{DEFA'B'C'} = \delta_A^{A'}\delta_{B}{}^{B'}\delta_{C}^{C'}$. This is reminiscent of the superpotential introduced in the TG theory. Indeed, compare the similarity between the last line of (\ref{superpotential}) and (\ref{DFTsuperpotential}).  The above equation coincides with the usual DFT equation of motion for the vielbein in the flux formulation as expected (see \cite{Geissbuhler:2013uka}).

\subsection{Examples of other connections in DFT}\label{DFTsection2exampotherconn} 
In this subsection we will find connections that furnish  non-trivial structures such as non-vanishing curvature and torsion and explicitly write them. They will transform under global (constant) double Lorentz transformations but the equations of motion derived from them restore the local double Lorentz transformations (much like in the teleparallel case). These connections will turn out to be characterized by different geometric structures compared to the generalized Levi-Civita and Weitzenb\"ock connection. 

We start by repeating a similar analysis as in subsection (\ref{eqforCoeff}) and see if we can find other connections determined in terms of the physical fields. 
The most general form of the connection $\Gamma$ that we are considering and consistent with its transformation properties under general coordinate transformations is:
\begin{equation}
\Gamma_{MN}{}^Q = W_{MN}{}^Q + E^{A}{}_ME^{B}{}_{N}E_{C}{}^Qw_{AB}{}^C,
\end{equation}
where the spin connection is given by:
\begin{align}\nonumber
	w_{A B}{}^C  = &  \,\, a_1\mathcal{F}_{A B}{}^C \,  + \, d_1 S_{A}{}^{A'} S_{B}{}^{B'} S^{C C'}\mathcal{F}_{A' B' C'} \\\nonumber 
	&  + \, b_1 S_{A}{}^{A'}\mathcal{F}_{A' B}{}^C + b_2 S_{B}{}^{B'}\mathcal{F}_{A B'}{}^C + b_3 S^{C C'}\mathcal{F}_{A B C'}\\\nonumber
	& + \, c_1 S_{A}{}^{A'}S_{B}{}^{B'}\mathcal{F}_{A' B'}{}^C + c_2 S_{A}{}^{A'}S^{C C'}\mathcal{F}_{A' B C'} + c_3 S_{B}{}^{B'} S^{C C'}\mathcal{F}_{A B' C'}\\\nonumber
	& + d_2\eta_{AB}\mathcal{F}{}^C + d_3\delta_{A}{}^C\mathcal{F}_{B} + d_4\delta_{B}{}^C\mathcal{F}_A\\\nonumber
	& + e_1\eta_{AB}S^{CC'}\mathcal{F}_{C'} + e_2\delta_{A}{}^C S_{B}{}^{B'}\mathcal{F}_{B'} + e_3\delta_{B}{}^C S_{A}{}^{A'}\mathcal{F}_{A'}\\\nonumber
	& + f_1 S_{AB}\mathcal{F}{}^C + f_2 S_{A}{}^C\mathcal{F}_{B} + f_3 S_{B}{}^C\mathcal{F}_A\\ \label{genericconDFT}
	& + g_1 S_{AB}S^{CC'}\mathcal{F}_{C'} + g_2 S_{A}{}^C S_{B}{}^{B'}\mathcal{F}_{B'} + g_3 S_{B}{}^C S_{A}{}^{A'}\mathcal{F}_{A'}.
\end{align}

At this point, the spin connection $w_{AB}{}^C$ does not transform well under local double Lorentz transformations but under constant ones. 

It is difficult to analyze generic cases for possible connections in DFT due to the amount of terms either in the connection (\ref{genericconDFT}) or in the generalized Ricci scalar. Therefore, we will restric to the particular case of connections satisfying 

\begin{equation}\label{eqrestexamofconn1}
w_{ABC} + w_{ACB} = 0,~~  
\end{equation}

\begin{equation}\label{eqrestexamofconn2}
S_{C}{}^{C'}w_{ABC'} + S_{B}{}^{B'}w_{ACB'} = 0.
\end{equation}

And, also, for simplicity, we add no dependence on $\mathcal{F}_{A}$ to the connection. This is because in the generalized Ricci scalar there could be no coupling between $\mathcal{F}_{A}$ and $\mathcal{F}_{ABC}$ due to the possible index contractions between them and $\eta^{AB}$ and $S^{AB}$. So we can, if we wish to, analyze terms with $\mathcal{F}_{A}$ and $\mathcal{F}_{ABC}$ in a separate way. The ansatz to the above requirements is then:
\begin{align}\nonumber
w_{A B}{}^C  = &  \,\, a_1\mathcal{F}_{A B}{}^C \,  + \, b_1 S_{A}{}^{A'} S_{B}{}^{B'} S^{C C'}\mathcal{F}_{A' B' C'} \\\nonumber 
&  + \, b_1 S_{A}{}^{A'}\mathcal{F}_{A' B}{}^C + b_2 S_{B}{}^{B'}\mathcal{F}_{A B'}{}^C + b_2 S^{C C'}\mathcal{F}_{A B C'}\\\nonumber
& + \, c_1 S_{A}{}^{A'}S_{B}{}^{B'}\mathcal{F}_{A' B'}{}^C + c_1 S_{A}{}^{A'}S^{C C'}\mathcal{F}_{A' B C'} + a_1 S_{B}{}^{B'} S^{C C'}\mathcal{F}_{A B' C'}\\
\end{align}

When plugging this ansatz into the generalized scalar curvature $\mathcal{R}(w)=P^{AB}P^{CD}\mathcal{R}_{ACBD}(w)$ we can extract the equations for the coefficients. We then have:

\begin{align}\nonumber
\mathcal{R}(w) = & A\, \mathcal{F}_{ABC}\mathcal{F}^{ABC} + B\, S^{AB} S^{CD} S^{EF} \mathcal{F}_{ACE} \mathcal{F}_{BDF} \,+\\
& C\,S^{AB} S^{CD} \mathcal{F}_{AC}{}^E \mathcal{F}_{BDE}\, + D\, S^{AB}\mathcal{F}_{A}{}^{CD}\mathcal{F}_{BCD}. 
\end{align}
Where 
\begin{align}\nonumber
A = &8b_{1} b_{1} + 8b_{1}b_{2} +8b_{2}b_{2} +8c_{1}c_{1} + 8a_{1}c_{1} - 4a_{1} \\
&- 16a_{1}b_{2} -16b_{1}c_{1} -8a_{1}b_{1} -8b_{2}c_{1} + 4b_{2}
\end{align}

\begin{align}\nonumber
B = & 16a_{1}b_{1} + 8b_{1}c_{1} + 8a_{1}b_{2} + 16b_{2}c_{1} - 4b_{1} -16a_{1}c_{1} \\
&-16b_{1}b_{2} - 4b_{1}b_{1} - 4b_{2}b_{2} - 4c_{1}c_{1} - 4a_{1}a_{1} + 4c_{1}
\end{align}

\begin{align}\nonumber
C = &40a_{1}c_{1} + 16a_{1}a_{1} + 16b_{1}b_{1} + 40b_{1}b_{2} + 16b_{2}b_{2}+ 16c_{1}c_{1} - 8c_{1} \\
& - 4a_{1} -40a_{1}b_{1} - 32a_{1}b_{2} - 32b_{1}c_{1} - 40b_{2}c_{1} + 8b_{1} + 4b_{2}
\end{align}

\begin{align}\nonumber
D = & 32a_{1}b_{1} + 40a_{1}b_{2} + 40b_{1}c_{1} + 32b_{2}c_{1} - 4b_{1} - 8b_{2}   \\\nonumber
& - 20a_{1}a_{1} - 32a_{1}c_{1}  - 20b_{1}b_{1} - 32b_{1}b_{2} - 20b_{2}b_{2} \\
&- 20c_{1}c_{1} + 8a_{1} + 4c_{1} + 8a_{1}a_{1}
\end{align}

By requiring that each term should vanish in order to get a vanishing generalized Ricci scalar we obtain the following solutions for the coefficients:

\begin{align}\label{dftexamplesotherconnect1}
a_{1}=0,~~b_{2}=-\frac{2}{3},~~c_{1}=\frac{1}{3}(-1 + 3b_{1}),\\\label{dftexamplesotherconnect2}
a_{1}=0,~~b_{2}=-\frac{1}{2},~~c_{1}=\frac{1}{2}(-1 + 2b_{1}),\\\label{dftexamplesotherconnect3}
a_{1}=0,~~b_{2}=-\frac{1}{6},~~c_{1}=\frac{1}{6}(1 + 6b_{1}),\\\label{eq4dftotherconnect}
a_{1}=0,~~b_{2}=0,~~c_{1}=b_{1}.
\end{align}

Now that we have found the coefficients  for the allowed connections, we compute several geometric quantities from them. The generalized torsion for them is given respectively by: 

\begin{align}\nonumber
\mathcal{T}^{(1)}_{MNP} & = (-7 + 12b_1)\mathcal{F}_{MNP} \, + \, (\frac{16}{3} - 16b_1)\big(\mathcal{F}_{M\dbar{N}P} + \mathcal{F}_{MN\dbar{P}} + \mathcal{F}_{\dbar{M}NP}\big)\\
&~~~~ +\,(-\frac{8}{3} + 20b_1)\big(\mathcal{F}_{\dbar{M}\dbar{N}P} + \mathcal{F}_{\dbar{M}N\dbar{P}} + \mathcal{F}_{M\dbar{N}\dbar{P}}\big) - \, 24b_1\mathcal{F}_{\dbar{M}\dbar{N}\dbar{P}}\,,\\[0.6cm]\nonumber
\mathcal{T}^{(1)}_{MNP} & = (-7 + 12b_1)\mathcal{F}_{MNP}\, + \,(6 - 16b_1)\big(\mathcal{F}_{M\dbar{N}P} + \mathcal{F}_{MN\dbar{P}} + \mathcal{F}_{\dbar{M}NP}\big)\\
&~~~~ +\,(-4 + 20b_1)\big(\mathcal{F}_{\dbar{M}\dbar{N}P} + \mathcal{F}_{\dbar{M}N\dbar{P}} + \mathcal{F}_{M\dbar{N}\dbar{P}}\big) - \, 24b_1\mathcal{F}_{\dbar{M}\dbar{N}\dbar{P}}\,,\\[0.6cm]\nonumber
\mathcal{T}^{(1)}_{MNP} & = (-7 + 12b_1)\mathcal{F}_{MNP}\, + \,(-\frac{2}{3} - 16b_1)\big(\mathcal{F}_{M\dbar{N}P} + \mathcal{F}_{MN\dbar{P}} + \mathcal{F}_{\dbar{M}NP}\big)\\
&~~~~ +\,(\frac{4}{3} + 20b_1)\big(\mathcal{F}_{\dbar{M}\dbar{N}P} + \mathcal{F}_{\dbar{M}N\dbar{P}} + \mathcal{F}_{M\dbar{N}\dbar{P}}\big) - \, 24b_1\mathcal{F}_{\dbar{M}\dbar{N}\dbar{P}}\,,\\[0.6cm]\nonumber
\mathcal{T}^{(1)}_{MNP} & = (-1 + 12b_1)\mathcal{F}_{MNP}\,  - 16b_1\big(\mathcal{F}_{M\dbar{N}P} + \mathcal{F}_{MN\dbar{P}} + \mathcal{F}_{\dbar{M}NP}\big)\\\label{dftgentorsion4sol}
&~~~~ +\, 20b_1\big(\mathcal{F}_{\dbar{M}\dbar{N}P} + \mathcal{F}_{\dbar{M}N\dbar{P}} + \mathcal{F}_{M\dbar{N}\dbar{P}}\big) - \, 24b_1\mathcal{F}_{\dbar{M}\dbar{N}\dbar{P}}\,.
\end{align}

It is  easy to show that $\nabla_M e^{-2d}=-e^{-2d}\mathcal{F}_M$ for the four connections (\ref{dftexamplesotherconnect1})-(\ref{eq4dftotherconnect}) found above. The solution (\ref{eq4dftotherconnect}) contains the Weitzenb\"ock connection by taking $b_1=0$ and indeed the generalized torsion reduces to (\ref{dftweitgtorsion}). Let us point out a couple of things. First of all, the generalized torsions have a very similar structures between each other, and, in particular, it can be noted that they all have common factors parametrized by $b_1$ which are  given by (\ref{dftgentorsion4sol}). This is because the three connections given by (\ref{dftexamplesotherconnect1})-(\ref{dftexamplesotherconnect3})  have a linear common dependence given by the connection (\ref{eq4dftotherconnect}). We have computed the generalized  curvature tensor for these connections and it is different from zero for all of them (its components are too long to be displayed), in this sense they are all different from the Weitzenb\"ock connection which furnishes a null curvature tensor. The generalized Ricci tensor is not that long, and, for completeness,  we display it in planar indices. Thus, the first term of (\ref{mastereqDFTforGamma}) for the four connections takes respectively the following form: 
\begin{align}\nonumber
 P^{C[B}\bar{P}^{A]D}\mathcal{R}_{C\dbar{E}D}{}^{\dbar{E}}(w) & =  \frac{8}{3}S_{F}{}^{C}S^{E[B}\mathcal{F}^{A]FD}\mathcal{F}_{ECD} - \frac{16}{3}S_{F}{}^{D'}S_{F'}{}^{E'}S^{C'[B}\mathcal{F}^{A]FF'}\mathcal{F}_{C'D'E'}\\\nonumber
& ~~~~ +  \frac{8}{3}S^{C[B}\mathcal{F}^{A]DE}\mathcal{F}_{CDE} - 4D^{C}\mathcal{F}^{AB}{}_{C} + 4D^{C'}\mathcal{F}^{ABD}S_{C'D}\\\label{dftgenriccitensorsol1}
&~~~~  - 4D
^{C'}\mathcal{F}^{E'FD'}S^{A}{}_{E'}S^{B}{}_FS_{C'D'} + 4 D^{C'}\mathcal{F}_{C'}{}^{D'E'}S^{A}{}_{D'}S^{B}{}_{E'} 
\end{align}

\begin{align}\nonumber
P^{C[B}\bar{P}^{A]D}\mathcal{R}_{C\dbar{E}D}{}^{\dbar{E}}(w) & = -4S^{FC'}S^{D'E'}S^{F'[B}\mathcal{F}^{A]}{}_{FD'}\mathcal{F}_{F'C'E'} 
-4S^{F[A}\mathcal{F}^{B]B'C'}\mathcal{F}_{FB'C'} \\\nonumber
&~~~~ - 4S_{FD'}D^{F}\mathcal{F}^{B'C'D'}S^{A}{}_{B'}S^{B}{}_{C'} + 4D
^{F}\mathcal{F}_{F}{}^{B'C'}S^{A}{}_{B'}S^{B}{}_{C'} \\\label{dftgenriccitensorsol2}
& ~~~~ - 4 D^{C}\mathcal{F}^{AB}{}_{C} + 4D^{C'}\mathcal{F}^{ABD}S_{C'D}
\end{align}

\begin{align}\nonumber
P^{C[B}\bar{P}^{A]D}\mathcal{R}_{C\dbar{E}D}{}^{\dbar{E}}(w) & = -\frac{8}{3}S^{E'C'}S^{F[B}\mathcal{F}^{A]}{}_{E'}{}^{D'}\mathcal{F}_{FC'D'} 
+ \frac{4}{3}S^{B'C'}S^{D'E'}S^{F[B}\mathcal{F}^{A]}{}_{B'D'}\mathcal{F}_{FC'E'} \\\label{dftgenriccitensorsol3}
&~~~~ -\frac{4}{3}S^{E'[A}\mathcal{F}^{B]D'C'}\mathcal{F}_{E'D'C'}
\end{align}

\begin{equation}\label{dftgenriccitensorsol4}
P^{C[B}\bar{P}^{A]D}\mathcal{R}_{C\dbar{E}D}{}^{\dbar{E}}(w)=0.
\end{equation}

We see that the parameter $b_1$ has dropped out from (\ref{dftgenriccitensorsol1})-(\ref{dftgenriccitensorsol4}). Although we have not displayed the generalized curvature tensor, however, this parameter does not drop out from it for any of the above connections. This means that these connections (and in particular solution  (\ref{eq4dftotherconnect}) with $b_1\neq 0$) are different from the Weitzenb\"ock connection from a geometric point of view\footnote{One may ask if the Bianchi identities $0=D_{[A}\mathcal{F}_{BCD]} - \frac{3}{4}\mathcal{F}_{[AB}{}^{E}\mathcal{F}_{CD]E}$ and $0=D^{C}\mathcal{F}_{CAB} + 2D_{[A}\mathcal{F}_{B]} - \mathcal{F}^{C}\mathcal{F}_{CAB}$ can be used to render the generalized curvature tensor and (or) the generalized Ricci tensor null. But a careful inspection shows that this is not the case.}. Interesting enough, however, connection (\ref{eq4dftotherconnect}) renders $P^{Q[N}\bar{P}^{M]R}\mathcal{R}_{Q\dbar{P}R}{}^{\dbar{P}}(\Gamma)$ equal to zero (\ref{dftgenriccitensorsol4}). We have seen a similar effect before commented in  subsection (\ref{eqforCoeff}). There we stated that there are gauge transformations for the curvature tensor and scalar curvature. We do not know if the generalized curvature tensor possesses any kind of gauge transformation as in equation  (\ref{gaugetransconncurvature}). If we compare the Riemannian curvature tensor in planar indices given by (\ref{RiemannFlat}) with the one in DFT given by (\ref{dftgenriemanncurvplanar}) (both for generic connections) we see that there are two important differences: The first one is that the usual Riemann curvature is antisymmetric in two of its indices, while the DFT Riemann tensor is symmetric under the exchange of its pairs of indices, i.e., $\mathcal{R}_{MNKL}=\mathcal{R}_{KLMN}$ (only when $\eta$-compatibility is enforced the share the same index properties). The second difference is that the generalized curvature tensor in planar indices makes use of the Weitzenb\"ock connection explicitly through the appearance of $\mathcal{F}_{ABC}=-3W_{[ABC]}$. Perhaps some sort of gauge transformation would necessarily include the use of $\mathcal{F}_{ABC}$ explicitly. The free parameter $b_1$ may be an indication that either a gauge transformation at the generalized curvature tensor level exists or  it is something particular of the generalized Ricci tensor and/or generalized scalar. In either case, the generalized Riemann curvature for the connections given by (\ref{dftexamplesotherconnect1})-(\ref{eq4dftotherconnect}) are different from the curvatures given by the Weitzenb\"ock connection. Thus, these connections are not gauge-artifacts of the generalized Weitzenb\"ock connection.

 From a geometric point of view, the first three connections given by the solutions (\ref{dftexamplesotherconnect1})-(\ref{dftexamplesotherconnect3})  yield equations of motion (for the generalized vielbein and dilaton) given by (\ref{mastereqDFTforGamma})-(\ref{mastereqdftdilaton}) and are represented geometrically by a non-vanishing generalized Ricci tensor, generalized contorsion tensor and L-tensor (through $\nabla e^{-2d}\neq 0$). The connection (\ref{eq4dftotherconnect}) yields only non-vanishing contorsion tensor and L-tensor. As a remark, the generalized torsion of all of these connections  turns out to be an antisymmetric tensor due to the condition $w_{ABC}=-w_{ACB}$. This implies that the expressions for components of the $\Sigma$ tensor are $\Sigma_{\dbar{M}\dbar{N}\dbar{P}}=\frac{1}{3}\mathcal{T}^{(1)}_{\dbar{M}\dbar{N}\dbar{P}}$ and $\Sigma_{\bar{M}\bar{N}\bar{P}}=\frac{1}{3}\mathcal{T}^{(1)}_{\bar{M}\bar{N}\bar{P}}$.  This means that not only the connections are fully determined but also provides a well-defined contorsion tensor (\ref{DFTconttensor}). We want to stress again that these connections, which we have dubbed them deformed versions of the Weitzenb\"ock connection, are well-behaved under general coordinate transformations (i.e they are a generalized scalar), and transform under constant double Lorentz transformations. When plugging these connections in the equations of motion (\ref{mastereqDFTforGamma}) and (\ref{mastereqdftdilaton}) we recover the local double Lorentz transformations.

\section{Summary}

\label{conclusions} \bigskip

We have expanded on geometric notions, such as contorsion tensor and non-metricity tensors in DFT. This has also allowed us to obtain other connections in DFT, besides the generalized Levi-Civita and Weitzenb\"ock connections, capable of describing the same dynamics of usual DFT in terms of other geometric quantities. The interpretation we are giving is that the new connections are deformed versions of the Weitzenb\"ock connection. We have first analyzed conditions on a generic connection (\ref{GammaConnectioncurvedsix}) in standard geometry in order to reproduce equivalent formulations to Einstein Gravity. We have done this by extracting information from the vanishing of the scalar curvature based on a general connection written in terms of first-order derivatives of the vielbein. By doing this, we have found in section (\ref{eqforCoeffMCcase}) that there are only four metric-compatible connections that satisfy the conditions, one of them being the Weitzenb\"ock connection. In this sense we are obtaining Teleparallel Gravity as a particular case. For a generic case that includes both torsion and non-metricity, there seems to be an infinite amount of solutions. In section (\ref{eqforCoeffNMcase}) we analyzed a particular case of a Weyl space with Weyl's vector being proportional to $\overset{(W)}{T}_{\mu}{}^{\sigma}{}_{\sigma}$ (the vector part of the Weitzenb\"ock torsion). The connections (\ref{gammaweylspace}) have torsion and non-metricity and are related to the metric-compatible ones (\ref{QzeroSolOneConn})-(\ref{QzeroSolFourConn}) through the transformation (\ref{gaugetransfRicciScalar}).
Our generic connections (except for the Weitzenb\"ock one) present non-vanishing curvature, torsion and non-metricity. The dynamics of the theories, defined by these connections, is represented by a mixture of those quantities. This can be seen in the equations of motion (\ref{eomgencasecurvedindices}) (or (\ref{eomgenericvarflat})) where the Ricci tensor, torsion and non-metricity enters in a non-trivial way.
  
With respect to the DFT part, we have followed the same procedure used in the GR case. We have found that it is possible to consider a decomposition of a generic DFT connection (equation (\ref{DFTfullconnectionLCContL})) in terms of a generalized Levi-Civita, generalized contorsion tensor and a L-tensor. Although this connection contains  seeming undetermined parts $\tilde{\Gamma}$ and $\Sigma$, we have seen that they can be determined. Indeed, in (\ref{DFTsection2undetpartconn}) we saw that it is possible to determine the undetermined part $\tilde{\Gamma}$ of the generalized Levi-Civita connection by requiring to be written as first-order derivatives acting on the generalized vielbein rather than $P$ or $\bar{P}$. And we have seen examples of how to determine the $\Sigma$-tensor in the Teleparallel equivalent case (subsection \ref{DFTsection2TDFT}) and in other examples of connections in subsection (\ref{DFTsection2exampotherconn})). We have found in subsection \ref{DFTsection2exampotherconn})  connections described by generic geometric quantities such as generalized torsion, $\nabla e^{-2d}$ and curvature, capable of describing the same DFT dynamics through the general equations of motion  (\ref{mastereqDFTforGamma}) and (\ref{mastereqdftdilaton}). These equations allow one to formulate the DFT dynamics in terms of other geometric quantities rather than the vanishing of the generalized Ricci tensor and generalized  Ricci scalar as in the standard approach to DFT using the generalized Levi-Civita connection. This means, in a sense, that it is not clear what is the fundamental connection to use in DFT. An observation is that we do not know if the generalized  curvature tensor possesses any kind of gauge transformation like in standard geometry (equation  (\ref{gaugetransconncurvature})). If we take a look at the solutions give in (\ref{dftexamplesotherconnect1})-(\ref{eq4dftotherconnect}) we saw that a free parameter $b_1$ has remained. Recall that the usual scalar curvature of Riemannian geometry has an invariance given by (\ref{gaugetransfRicciScalar}) which descends from a generalization of \ref{gaugetransconncurvature}) (see also the discussion in subsection (\ref{GRgaugeredundancysection})). The free parameter $b_{1}$ might be  an indication that some sort of gauge transformation for the generalized curvature tensor could exist. However, due to the differences in the geometric quantities characterizing  our deformed versions of the Weitzenb\"ock connection,  we concluded that these are not gauge-artifacts of the generalized Weitzenb\"ock connection. We content ourselves with showing, as we did above, that non-trivial solutions, other than the generalized Levi-Civita connection and the generalized Weitzenb\"ock connection, exist and yield an equivalent theory (in terms of dynamical degrees of freedom) to DFT and yet described by different geometric quantities. It is important to stress again, that this equivalence with the DFT dynamics is due to the same global-local mechanism of the TG case either in GR or DFT.

As a future work, there are a couple of things to explore. It might be useful to understand if the teleparallel equivalent of DFT can be interpreted as a gauge theory, in a similar fashion as in the teleparallel equivalent of GR. There, TEGR can be understood as arising from gauging the translation group $\mathbb{R}^4$ where the gauge potential is related to the vielbein (see for instance \cite{Arcos:2005ec} for reviews). In DFT, we might try to follow a similar procedure keeping in mind the periodicity of the doubled-torus bundle and the section condition, so in this case a compact subspace of $\mathbb{R}^{2D}$ would be gauged. Another line of research would be to explore applications to $\alpha'$-corrections. As said in the Introduction, higher-order derivatives terms in the DFT action are important to understand how T-duality constraints or gives information about possible higher-order curvature completion to an effective action coming from string theory. This work has shown that it is possible to rewrite DFT in terms of other geometric quantities not based only on generalized curvatures as in standard DFT. For instance, any of the connections found in subsection(\ref{DFTsection2exampotherconn}), including the Weitzenb\"ock connection, are described, generically, by a curvature tensor, generalized contorsion and L-tensor through  $\nabla e^{-2d}$. In particular, the Weitzenb\"ock case, gives the teleparallel equivalent of DFT and it is only described by a generalized contorsion tensor and non-vanishing $\nabla e^{-2d}$. It might well be possible to describe higher-order derivative terms in the teleparallel equivalent coming from these quantities, which is another motivation for this work: $\alpha'$-corrections in DFT \cite{Hohm:2013jaa,Marques:2015vua,Coimbra:2014qaa,Lee:2015kba}. In general it would be useful to have all of the components of the generalized Riemann curvature to be well-defined in order to write down higher-order derivative corrections to the effective action. As said before, when writing in terms of generalized metric or dilaton, the generalized Riemann curvature cannot be fully determined in terms of them. This suggests $\alpha'$ corrections to generalized diffeomorphisms should be taken into account to have and $\alpha'$ corrections to the effective action. Even if there were a physical generalized Riemann curvature built from the generalized metric and dilaton  it could not be of use in constructing a Riemann squared \cite{Hohm:2013jaa}. In a different approach \cite{Marques:2015vua}, the  frame-formalism of DFT is used to address the $\alpha'$-corrections. This formalism turns out to be slightly more flexible since it encompasses previous $\alpha'$-DFT proposals based on generalized metric formalism. It also maintains the usual form of generalized diffeomorphisms (no $\alpha'$-corrections) but the local double Lorentz transformations receive corrections plus an anomalous transformation of the generalized metric under them. These non-standard transformations determine the structure of the four-derivative corrections. Given the fact that the connections described in this paper are fully determined in terms of and built out from the generalized frame of DFT, it might have some relevance to the understanding of the $\alpha'$-corrections in DFT. For instance, the $f(T)$-theories mentioned in the Introduction are higher-order derivative corrections to the usual TG (or Einstein-Hilbert action). The connections to be described here contain, for instance, non-trivial generalized torsion $\mathcal{T}$, so any $f(\mathcal{T})$-theory to be considered will naturally contain higher-order corrections and may be a supplement for the standard approach of $\alpha'$-corrections in terms of Riemann powers. 

 In any case, important attention to the local double Lorentz transformations must be paid, since, a priori, it is not clear what combinations of higher-order terms coming from these geometric quantities will keep partially or not at all the local double Lorentz transformations. In our approach, when considering the connections as written in terms of first-order derivatives of the (generalized) vielbein they are fully determined in terms of the physical field and, also, they generally break the local (double) Lorentz transformations. Only the global group (i.e. with constant parameters) survives. However, as said before, our actions and equations of motion derived form these connections restore the local (double) Lorentz group and this might also apply to a $f(C)$-theory where $C$ includes torsion, non-metricity, etc.

These $f(\mathcal{T})$-theories are out of the scope of this work, but, as a first step, it is useful to know that they may be constructed using the connections described in this paper.

\section*{Acknowledgments}

The author thanks R. Ferraro and F. Fiorini for very useful discussions in Teleparallel Gravity. We thank A. Chatzistavrakidis, A. Goya and D. Marques for carefully reading this manuscript and suggestions and we also thank G. Aldazabal and Falk Hassler for some comments. Part of the calculations of this project have been done with Cadabra Software \cite{Peeters:2006kp}. This work is
partially supported by CONICET grant PIP-11220110100005 and PICT-2016-135.

\end{document}